\documentclass[a4paper,11pt]{article}

 \usepackage{amsfonts,amsthm,amssymb}
 \usepackage{latexsym, epsf}

\newcommand{\eq}{\begin{equation}}
\newcommand{\en}{\end{equation}}
\newcommand{\eqa}{\begin{eqnarray}}
\newcommand{\ena}{\end{eqnarray}}

\begin{document}

\setlength{\unitlength}{1mm}

\thispagestyle{empty}

\vspace*{0.1cm}

\begin{center}
{\LARGE \bf Yang--Baxterizations, Universal Quantum Gates and Hamiltonians  \\[2mm] }

\vspace{.5cm}

Yong Zhang\footnote{yong@itp.ac.cn}\\

${}$ Institute of Theoretical Physics, Chinese Academy of Sciences\\
P. O. Box 2735, Beijing 100080, P. R. China\\[0.4cm]

Louis H. Kauffman\footnote{kauffman@uic.edu}\\

Department of Mathematics, Statistics and Computer Science\\
University of Illinois at Chicago\\ 851 South Morgan Street,
Chicago, IL, 60607-7045\\[0.4cm]

Mo-Lin Ge\footnote{GEML@nankai.edu.cn}\\

${}$ Nankai Institute of Mathematics, Nankai University\\ Tianjin
300071, P. R. China\\[0.4cm]

\end{center}

\vspace{0.2cm}

\begin{center}
\parbox{12cm}{
\centerline{\small  \bf Abstract}
 \small \noindent
\\
The unitary braiding operators describing topological entanglements
can be viewed as universal quantum gates for quantum computation.
With the help of  the Brylinski's theorem, the unitary solutions of
the quantum Yang--Baxter equation can be also related to universal
quantum gates. This paper derives unitary solutions of the quantum
Yang--Baxter equation via Yang--Baxterization from the solutions of
the braid relation. We study Yang--Baxterizations of the
non-standard and standard representations of the six-vertex model
and the complete solutions of the non-vanishing eight-vertex model.
We construct Hamiltonians responsible for the time-evolution of the
unitary braiding operators which lead to the Schr{\"o}dinger
equations.
 }

\end{center}

\vspace*{10mm}
\begin{tabbing}

\\ Yang--Baxter, Universal quantum gates\\


\end{tabbing}

\newpage

\section{Introduction}

There are natural relationships between quantum entanglement
\cite{nielsen} and topological entanglement \cite{kauffman0}.
Topology studies global relationships in spaces, and how one space
can be placed within another, such as knotting and linking of curves
in three-dimensional space. One way to study topological
entanglement and quantum entanglement is to try making direct
correspondences between patterns of topological linking and
entangled quantum states. One approach of this kind was initiated by
Aravind \cite{aravind}, suggesting that observation of a link would
be modelled by deleting one component of the link. But this
correspondence property of quantum states and topological links is
not basis independent \cite{aravind}.

A deeper method (we believe) is to consider unitary gates $\check R$
that are both universal for quantum computation and are also
solutions to the condition for topological braiding. Such $\check
R$-matrices are unitary solutions to the Yang--Baxter equation (the
braid relation). We are then in a position to compare the
topological and quantum properties of these transformations. In this
way, we can explore the apparently complex relationship among
topological entanglement, quantum entanglement, and quantum
computational universality. This way has been explored in a series
of papers \cite{kauffman1, kauffman2, kauffman3, kauffman4,
 kauffman6, kauffman7,kauffman8, yong, dye}.

The present paper derives unitary solutions of the Quantum
Yang--Baxter Equation (QYBE) via Yang--Baxterization and explores
the corresponding dynamical evolution of quantum entanglement
states. The solutions to the QYBE that we derive by
Yang--Baxterization contain a spectral parameter $x$, and hence do
not, except in special cases, give representations of the Artin
braid group. These new solutions are unitary, and they do give
useful quantum gates. Thus we show in this paper that the full
physical subject of solutions to the quantum Yang--Baxter equation
(including the spectral parameter $x$) is of interest for quantum
computing and quantum information theory.

The plan of the paper is organized as follows. In the second
section, a unitary braiding operator is regarded as a quantum
entanglement operator and further a universal quantum gate. To
describe the dynamical evolution of the unitary braiding operator
$\check{R}$, Yang--Baxterization is used to solve the QYBE. In the
third section, the unitary solutions of the QYBE for the
non-standard and standard representations of the six-vertex model
are obtained via Yang--Baxterization. In the fourth section, the
complete unitary solutions of the QYBE for the non-vanishing
eight-vertex model are obtained via Yang--Baxterization. In the
fifth section, with the Brylinksi's theorem \cite{BB}, all unitary
$\check{R}(x)$-matrices presented in this paper are recognized to be
universal quantum gates for most $x$-values. In the sixth section,
the Hamiltonian determining the time evolution of quantum state is
constructed with the unitary $\check{R}(x)$-matrix. In the seventh
section, as an example, the CNOT gate is constructed in terms of the
unitary $\check{R}$-matrix ($\check{R}(x)$-matrix) and local unitary
transformations. To conclude, remarks on our work are made. In
Appendix A, a pragmatic introduction to Yang--Baxterization is
presented.

\section{The QYBE in quantum entanglements}

This section presents  basic elements underlying our work. The Braid
Group Representation (BGR) and the QYBE are described in the sense
of studying quantum entanglements. The Brylinski's theorem \cite{BB}
plays the key role in relating a unitary $\check{R}$-matrix
($\check{R}(x)$-matrix) to a universal quantum gate.
Yang--Baxterization derives the corresponding $\check{R}(x)$-matrix
from the $\check{R}$-matrix and makes it possible constructing the
Hamiltonian determining the evolution of the unitary braiding
operator ($\check{R}$-matrix).

\subsection{The BGR in quantum entanglements}

Braids are patterns of entangled strings. A braid has the form of
a collection of strings extending from one set of points to
another, with a constant number of points in each cross section.
Braids start in one row of points and end in another. As a result,
one can multiply two braids to form a third braid by attaching the
end points of the first braid to the initial points of the second
braid. Up to topological equivalence, this multiplication gives
rise to a group, the Artin braid group $B_{n}$ on $n$ strands,
which is generated by $ \lbrace b_i | 1 \leq i \leq n-1 \rbrace $.

The group $ B_n $ consists of all words of the form $ b_{j_1} ^{
\pm 1} b_{j_2} ^{ \pm 1} ... b_{j_n} ^{ \pm 1} $ modulo the
relations: \eqa
   b_{i}  b_{i+1} b_{i} &=& b_{i+1} b_{i} b_{i+1}, \qquad
   1 \leq i \leq n-1, \nonumber\\
   b_{i} b_{j} &=& b_{j} b_{i}, \qquad
   | i - j | > 1.
\ena Each braid is, in itself, a pattern of entanglement. Each
braid is also an operator that operates on other patterns of
entanglement (braids) to produce new entanglements (braids again).

The analogy between topological entanglement and quantum
entanglement (from the point of view of braids) means {\em the
association of a unitary operator with a braid that respects the
topological structure of the braid and allows exploration of the
entanglement properties of the operator.} In other words, we propose
to study the analogy between topological entanglement and quantum
entanglement by looking at {\em unitary representations of the Artin
braid group}. The main point for the exploration of the analogy is
that (from the point of view of a BGR) each braid is seen as an
operator rather than a state. See Fig. 1.
\begin{figure}[!hbp]
\begin{center}
\epsfxsize=12.5cm \epsffile{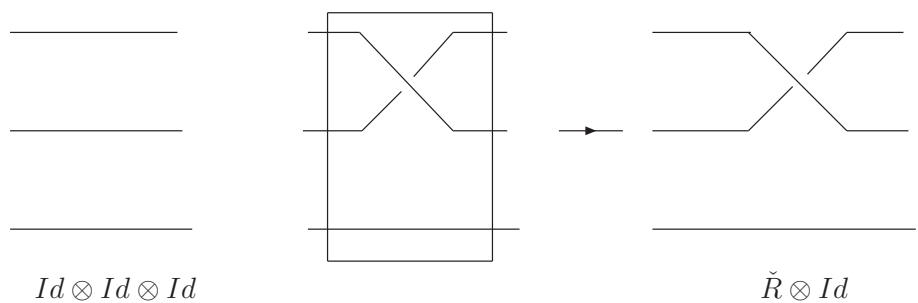} \caption{A braiding
operator $\check{R}\otimes Id$. } \label{fig1}
\end{center}
\end{figure}

Consider representations of the Artin braid group constructed in the
following manner. To an elementary two strand braid there is
associated an operator \eq \check{R}: V \otimes V \longrightarrow V
\otimes V.\en Here $V$ is a complex vector space, and for our
purposes, $V$ will be two dimensional so that $V$ can hold a single
qubit of information. One should think of the two input and two
output lines from the braid as representing this map of tensor
products. Thus the left endpoints of $\check{R}$ as shown in Fig. 1
and Fig. 2 represent the tensor product $V \otimes V$ that forms the
domain of $\check{R}$ and the right endpoints of the diagram for
$\check{R}$ represent $V \otimes V$ as the range of the mapping. In
the diagrams with three lines shown in Fig. 2, we have mappings from
$V \otimes V \otimes V$ to itself.
\begin{figure}
\begin{center}
\epsfxsize=14.cm \epsffile{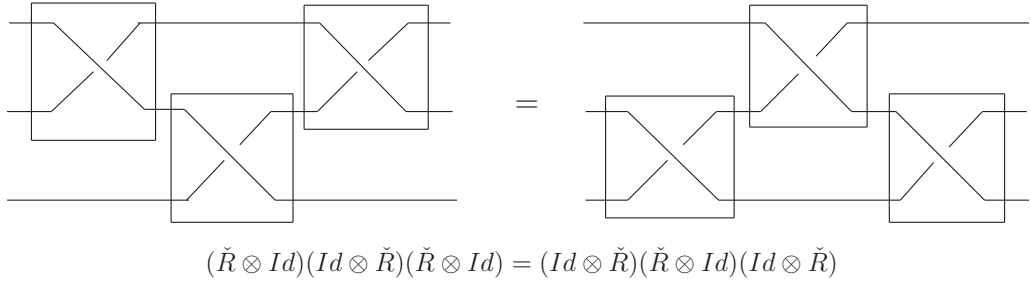} \caption{The Yang--Baxter
equation (the braid relation).} \label{fig2}
\end{center}
\end{figure}

The identity shown in Fig. 2 is called the Yang--Baxter Equation
(the braid relation), and it reads algebraically as follows, where
$Id$ denotes the identity transformation on $V.$ \eq \label{ybe}
(\check{R} \otimes Id)(Id \otimes \check{R})(\check{R} \otimes Id) =
(Id \otimes \check{R})(\check{R} \otimes Id)(Id \otimes \check{R}).
\en This equation expresses the fundamental topological relation in
the Artin braid group, and is the main requirement for producing a
representation of the braid group by this method.

\subsection{The Brylinski's theorem and $\check{R}$-matrix}

A two-qubit gate $G$ is a unitary  linear mapping $G:V \otimes V
\longrightarrow V \otimes V$ where $V$ is a two complex dimensional
vector space. A gate $G$ is said to be {\it entangling} if there is
a vector $$| \alpha \beta \rangle = | \alpha \rangle \otimes | \beta
\rangle \in V \otimes V$$ such that $G | \alpha \beta \rangle$ is
not decomposable as a tensor product of two qubits. Under these
circumstances, one says that $G | \alpha \beta \rangle$ is
entangled.

In \cite{BB}, the Brylinskis give a general criterion of $G$ to be
universal  (in the presence of local unitary transformations).
They prove that a two-qubit gate $G$ is universal if and only if
it is {\em entangling}.

Here is the specific $\check{R}$-matrix that we shall examine
\eq \check{R} = \left( \begin{array}{cccc} a & 0 & 0 & 0 \\ 0 & 0 & d & 0 \\
0 & c & 0 & 0 \\ 0 & 0 & 0 & b \end{array} \right)\en where
$a,b,c,d$ can be any scalars on the unit circle in the complex
plane. Then $\check{R}$ is a unitary matrix and it is a solution
to the Yang--Baxter equation (\ref{ybe}).

The point of this case study is that $\check{R}$, being unitary, can
be considered as a universal quantum gate and since $\check{R}$ is
the key ingredient in a unitary representation of the braid group,
it can be considered as an operator that performs topological
entanglement. We shall see that it can also perform quantum
entanglement in its action on quantum states. The $\check{R}$-matrix
can also be used to make an invariant of knots and links that is
sensitive to linking numbers \cite{kauffman0}.

\begin{figure}[!hbp]
\begin{center}
\epsfxsize=8.cm \epsffile{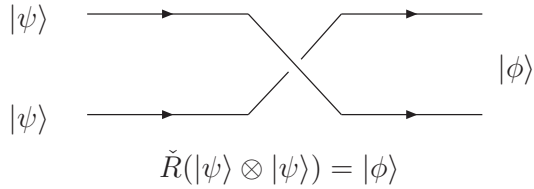} \caption{The braiding
operator $\check{R}$ as an entangling operator.} \label{fig3}
\end{center}
\end{figure}

Consider the action of the unitary transformation $\check{R}$ on
quantum states. We have \eqa \check{R}|00\rangle &=& a|00\rangle,
\qquad \check{R}|01\rangle = c|10\rangle, \nonumber\\
\check{R}|10\rangle &=& d|01\rangle, \qquad \check{R}|11\rangle =
b|11\rangle. \ena
 Here is an elementary proof that the operator
$\check{R}$ can entangle quantum states. If $\check{R}$ is chosen so
that $ab \ne cd$, then the state $\check{R}(|\psi\rangle \otimes
|\psi\rangle)$, with $|\psi\rangle = |0\rangle + |1\rangle$, is
entangled as a quantum state since \eq
|\phi\rangle=\check{R}(|\psi\rangle \otimes |\psi\rangle) =
a|00\rangle + c|10\rangle + d|01\rangle + b|11\rangle\en for $ab \ne
cd$ (see Fig. 3) is an entangled state.

 \subsection{Yang--Baxterization and Hamiltonian }

In this paper, the BGR $b$-matrix  \cite{kauffman2} and the QYBE
solution $\check{R}$-matrix \cite{yang, baxter, faddeev, jimbo}
are $n^2 \times n^2$ matrices acting on $ V\otimes V$ where $V$ is
an  $n$-dimensional vector space. As $b$ and $\check{R}$ act on
the tensor product $V_i\otimes V_{i+1}$, we denote them by $b_i$
and  $\check{R}_i$, respectively.

 The BGR $b$-matrix has to satisfy the braid  relation \eq \label{bgr}
 b_i\,b_{i+1}\,b_i=b_{i+1}\,b_i\,b_{i+1}, \en while the QYBE has
 the form \eq \label{qybe}
 \check{R}_i(x)\,\check{R}_{i+1}(xy)\,\check{R}_i(y)=
 \check{R}_{i+1}(y)\,\check{R}_i(xy)\,\check{R}_{i+1}(x)\en with the
 asymptotic condition \eq \check{R}(x=0)=b, \en and $x$ called the
 spectral parameter. From these two equations both $b$ and
 $\check{R}(x)$ are fixed up to an overall scalar factor.

In terms of the permutation operator $P$ specified by $P(\xi
 \otimes \eta )=\eta\otimes \xi$ and the $\check{R}(x)$-matrix, the
 solution of the algebraic QYBE reading \eq
 R_{12}(x)R_{13}(xy)R_{23}(y)=R_{23}(y)R_{13}(xy)R_{12}(x) \en
 takes the form \eq R(x)=P\check{R}(x) \en where $R_{ij}$ is an
 operator acting on the tensor product $V_i \otimes V_j$. But we will
 mainly deal with the unitary $4\times 4$ $\check{R}(x)$-matrix since
 the unitary permutation operator $P$ can be found in order to
 obtain the unitary $R(x)$-matrix.

  As a solution of the QYBE (\ref{qybe}), the $\check{R}(x)$-matrix usually depends
  on a deformation parameter $q$ and the spectral parameter $x$. With two such parameters,
  there exist two approaches to solving the QYBE (\ref{qybe}). Taking the limit as $q\to 1$
  leads to the classical $r$-matrix satisfying the classical Yang--Baxter equation.
  Then $q$-deforming $r$-matrices as solutions of the classical Yang--Baxter equation
  regain the q-dependence for the $\check{R}$-matrices as solutions for the QYBE (\ref{qybe}).
  The QYBE solution via this line of thought has been systematized by a general strategy in
  quantum groups (Hopf algebras) approach
   \cite{drinfeld,jimbo1,drinfeld1,reshetikhin1,reshetikhin2, jimbo2}.

 Taking the limit as $x\to 0$ leads to the braid relation
 (\ref{bgr}) from
  the QYBE (\ref{qybe}) and the BGR $b$-matrix from the $\check{R}(x)$-matrix. Concerning relations
  between the BGR and $x$-dependent solutions of the QYBE (\ref{qybe}), we either reduce a known
  $\check{R}(x)$-matrix to the BGR $b$-matrix, see \cite{wadati1, wadati2, turaev}, or construct
  the $\check{R}(x)$-matrix from a given BGR $b$-matrix. Such a construction is called
  Yang--Baxterization.

  In knot theory, these solutions were first studied by Jones
  \cite{jones} and Turaev \cite{turaev} for the BGR (\ref{bgr}) satisfying
  the Hecke algebra relations and for the BGR (\ref{bgr}) satisfying the
  Birman--Wenzl algebra relations (corresponding to the Kauffman two-variable
  polynomial, see \cite{kauffman0}). Later the more general cases with a BGR (\ref{bgr}) having
  three or four unequal eigenvalues were considered \cite{molin1, molin2, molin4},
  including all known trigonometric solutions to the QYBE (\ref{qybe}). Also
  Yang--Baxterization of non-standard BGR $b$-matrix has been
  discussed \cite{sogo,molin6, molin7, molin8, molin9, lee1, lee2, schultz1}.

  In this paper, via Yang--Baxterization the unitary
  $\check{R}(x)$-matrices are derived and used to construct
  Hamiltonians determining the time evolution of quantum
  states. In Appendix A, a pragmatic revisit to
  Yang--Baxterization has been given.

 \section{Unitary $\check R(x)$-matrix: the six-vertex model}

 In this section, we deal with Yang--Baxterizations of both the
 standard representation and the non-standard representation of
 the six-vertex model in detail,  as an example of
 deriving the unitary $\check{R}(x)$-matrix with the spectral
 parameter $x$ from the given BGR $b$-matrix.

 Consider a non-standard BGR $b$-matrix suitable for constructing the
 Alexander polynomial \cite{kauffman0} \eq
 b=\left(\begin{array}{cccc}
 q & 0 & 0 & 0 \\
 0 & 0 & 1 & 0 \\
 0 & 1 & q-q^{-1} & 0 \\
 0 & 0 & 0 & -q^{-1}
 \end{array}\right)
 \en  where the deformation parameter $q$ has been assumed to be
non-vanishing. Let $b^\dag$  the transpose and conjugation of $b$.
The unitarity condition $b^\dag b= b^\dag b =1\!\!1$ leads to $q=\pm
1$.

It has two distinct eigenvalues: $q$ and $-q^{-1}$. By
Yang--Baxterization as in Appendix A.1, the BGR $b$-matrix
corresponds to the $\check{R}(x)$-matrix \eq \label{rmatrixone}
\check{R}(x)=\left(\begin{array}{cccc}
q-q^{-1} x & 0 & 0 & 0 \\
0 & (q-q^{-1}) x & 1-x & 0 \\
0 & 1-x & q-q^{-1} & 0 \\
0 & 0 & 0 & q x-q^{-1}
\end{array}\right)
\en which satisfies the QYBE (\ref{qybe}). The case of $x=1$ will
not be considered since $\check{R}(1)=(q-q^{-1})\,1\!\! 1$.

Assume that $q, x$ are complex numbers, $x \neq 1$ and $q \neq 0$,
with the complex conjugations $\bar q,\bar x$ and the norms
$\|q\|^2:=q \bar q$, $\|x\|^2:=x \bar x$. The $\check{R}(x)$-matrix
has its conjugate matrix by\eq \check{R}^\dag(\bar
x)=\left(\begin{array}{cccc}
\bar q- {\bar q}^{-1} \bar x & 0 & 0 & 0 \\
0 & (\bar q-{\bar q}^{-1}) \bar x & 1-\bar x & 0 \\
0 & 1-\bar x & \bar q-{\bar q}^{-1} & 0 \\
0 & 0 & 0 & \bar q \bar x-{\bar q}^{-1}
\end{array}\right).
\en The unitarity condition shows
 \eq  \label{unitarity} \check{R}(x)\check{R}^\dag(\bar x)=\check{R}^\dag(\bar x)
  \check{R}(x)=\rho 1\!\!1  \en
 where $\rho$ is a normalization factor of the $\check{R}(x)$-matrix. It gives us
 the following equations \eq \left\{\begin{array}{ccc}\|q-q^{-1}x\|^2=\|q x-q^{-1}\|^2
 &=& \rho\\ (1-x)(\bar q-\bar q^{-1}) \bar x+(q-q^{-1})(1-\bar x)&=& 0
 \\ \|q-q^{-1}\|^2\,\, \|x\|^2+\|1-x\|^2&=& \rho\\
 \|q-q^{-1}\|^2 +\|1-x\|^2&=& \rho \end{array} \right.. \en

  Comparing the last two equations, we obtain \eq \frac 1 {\|q\|^2}\,\|1-q^2\|^2
  (1-\|x\|^2)=0 \en which leads to the first
  case of $q=\pm 1$ and the second case of  $\|x\|=1$. With
  $\|x\|=1$ and \eq q^{-1}=\frac {\bar q} {\|q\|^2}, \qquad \bar q^{-1}=\frac {q}
  {\|q\|^2}, \en going further to rewrite the above second equation
 as \eq (1-\bar x)(1+\frac 1 {\|q\|^2})(q-\bar q)=0 \en which shows
 that $q$ has to be real. Analyzing the above first equation gives
 \eq \frac 1 {\|q\|^2}\,(q+\bar q)(q-\bar q)(x-\bar x)=0 \en which is
 satisfied for real $q$. To make a safe judgement, comparing the first
 equation and the last two equations, we have \eq
 (q-\bar q)(q(1-\bar x)+\bar q (1-x))=0,
 \en
 which is also satisfied for real $q$. Therefore, as $\|x\|=1, x \neq 1$
 and $q$ real, the $\check{R}(x)$-matrix obtained through Yang--Baxterization is
 unitary.

In the first case that $q=\pm 1, x\neq 1$, we have \eq \rho=\| 1-x
\|^2:=(1-x)(1-\bar x), \en and then the corresponding
$\check{R}(x)$-matrix is given by \eq \check{R}(x)|_{q=\pm
1}=\frac {1-x} {\| 1-x \|}\,\,b|_{q=\pm 1}. \en  In the second
case, taking $x=e^{2\,i\theta}, \theta \neq 0$ and $q=e^\gamma$,
the $\check{R}(x)$-matrix takes the form \eq \label{rrmatrixone}
\check{R}(\theta)=2\,e^{i\theta} \left(\begin{array}{cccc}
\sinh(\gamma-i\theta)& 0 & 0 & 0 \\
0 &  e^{i\theta}\sinh\gamma & -i\sin{\theta} & 0 \\
0 & -i\sin{\theta} & e^{-i\theta}\sinh\gamma  & 0 \\
0 & 0 & 0 & \sinh(\gamma+i\theta)
\end{array}\right).
\en Modulo the scalar factor $2 e^{i\theta}$, the unitary
$\check{R}(\theta)$-matrix has the normalization factor $\rho$ by
\eq \label{ucsix} \rho=\sinh^2\gamma+\sin^2\theta. \en In
addition, with the new variable $\theta$, such
$\check{R}(\theta)$-matrix satisfies the following QYBE: \eq
\check{R}_{12}(\theta_1)\,\check{R}_{23}(\theta_1+\theta_2)\,\check{R}_{12}(\theta_2)=
\check{R}_{23}(\theta_1)\,\check{R}_{12}(\theta_1+\theta_2)\,\check{R}_{23}(\theta_2).
\en Similarly, we can treat the case of $q=- e^\gamma$.

Let consider the BGR $b$-matrix which satisfies the Temperley--Lieb
algebra and plays the role in constructing the Jones polynomial, see
\cite{kauffman1}, \eq b=\left(\begin{array}{cccc}
q & 0 & 0 & 0 \\
0 & 0 & 1 & 0 \\
0 & 1 & q-q^{-1} & 0 \\
0 & 0 & 0 & q
\end{array}\right).
\en It has two eigenvalues: $q$ and $-q^{-1}$. With
Yang--Baxterization, the corresponding $\check{R}(x)$-matrix has
the form \eq
\label{rmatrixtwo}\check{R}(x)=\left(\begin{array}{cccc}
q-q^{-1} x & 0 & 0 & 0 \\
0 & (q-q^{-1}) x & 1-x & 0 \\
0 & 1-x & q-q^{-1} & 0 \\
0 & 0 & 0 & q-q^{-1} x
\end{array}\right).
\en  Similar to the preceding case for the Alexander matrix, the
unitarity condition (\ref{unitarity}) informs us that the spectral
parameter $x$ lives on the unit circle $\|x\|=1$ and the
deformation parameter $q$ has to be real.

We apply the ordinary unitarity requirements (\ref{unitarity})
instead of the unitarity conditions $\check R(x^{-1}) \check
R(x)\propto 1\!\! 1$ often mentioned in literature related to the
QYBE (\ref{qybe}). With the latter one, we have the following the
normalization factor $\rho$ by
 \eq
 \rho=\check R(x)\check{R}(x^{-1})=(q^2+q^{-2}-x-x^{-1})
 \en
for the non-standard representation and standard representation of
the six-vertex model. It is compatible with the normalization
factor (\ref{ucsix}).

\section{Unitary $\check R(x)$-matrix: the eight-vertex model}

 In this section, we will present the complete solutions of the
 BGR (\ref{bgr}) for the non-vanishing eight-vertex model and the corresponding unitary
 $\check{R}(x)$-matrices via Yang--Baxterization. In terms of non-vanishing
 Boltzman weights $w_1$, $w_2$, $\cdots$, $w_8$, the BGR $b$-matrix of the eight-vertex
 model assumes the form
 \eq
 b=\left(\begin{array}{cccc}
 w_1 & 0 & 0 & w_7 \\
 0 & w_5 & w_3 & 0 \\
 0 & w_4 & w_6 & 0 \\
 w_8 & 0 & 0 & w_2
 \end{array}\right).\en
Choosing suitable Boltzman weights leads to unitary solutions of the
Yang--Baxter equation or the braid relation (\ref{bgr}). In Appendix
A.2, Yang--Baxterization used here is sketched in a practical way.

\subsection{Solutions of the eight-vertex model for $w_3=-w_4$ (I)}

Setting $w_5=w_1=w_2=w_6$ gives us $w_1^2=w_3^2=w_4^2$ and
$w_3^2+w_7 w_8=0$, see Appendix A.2. In the case of $w_3\neq w_4$,
we have $w_3=-w_4$ and $w_1=\pm w_3$. The BGR $b$-matrix has the
form \eq b_\pm=\left(\begin{array}{cccc}
w_1 & 0 & 0 & w_7 \\
0 & w_1 & \pm w_1 & 0 \\
0 & \mp w_1 & w_1 & 0 \\
-\frac {w_1^2} {w_7} & 0 & 0 & w_1
\end{array}\right)  \Longleftrightarrow
\left(\begin{array}{cccc}
1 & 0 & 0 & q \\
0 & 1 & \pm 1 & 0 \\
0 & \mp 1 & 1 & 0 \\
-q^{-1} & 0 & 0 & 1
\end{array}\right).\en

It has two eigenvalues $\lambda_1=1-i$ and $\lambda_2=1+i$.  The
corresponding $\check{R}(x)$-matrix via Yang--Baxterization is
obtained to be \eqa  \label{belltype1} \check{R}_\pm(x) &=&
 b_\pm + x\,\lambda_1\lambda_2\, b_\pm^{-1}  \nonumber\\
&=& \left(\begin{array}{cccc}
1+x & 0 & 0 & q(1-x) \\
0 & 1+x & \pm(1-x) & 0 \\
0 & \mp(1-x) & 1+x & 0 \\
-q^{-1} (1-x) & 0 & 0 & 1+x \end{array}\right).  \ena

Assume the spectral parameter $x$ and the deformation parameter
$q$ to be complex numbers. The unitarity condition
(\ref{unitarity}) leads to the following equations
 \eq \left\{ \begin{array}{ccc}
\|1+x\|^2+\|q\|^2 \|1-x\|^2 &=& \rho \\
\|1+x\|^2+\frac 1 {\|q\|^2} \|1-x\|^2 &=& \rho\\
\|1+x\|^2+\|1-x\|^2 &=& \rho\\
(1-x)(1+\bar x )-(1+x) (1-\bar x)&=& 0\\
-q^{-1} (1-x)(1+\bar x )+\bar q\,(1+x) (1-\bar x)&=&
0\end{array}\right. \en which specify $x$ real and $q$ living at a
unit circle.

Introducing the new variables of angles $\theta$ and $\varphi$ as
follows \eq \label{belltypetran} \cos\theta=\frac 1
{\sqrt{1+x^2}}, \qquad
 \sin\theta=\frac x {\sqrt{1+x^2}}, \qquad
 q=e^{-i\varphi},
 \en
we represent the $\check{R}_\pm(x)$-matrix in a new form \eq
\label{bellrmatrix} \check{R}_\pm(\theta)=\cos\theta\,\,
b_\pm(\varphi)+\sin\theta\,\, (b_\pm)^{-1}(\varphi)\en in which
the BGR $b_\pm(\varphi)$-matrices are given by \eq \label{bgrphi}
 b_\pm(\varphi)=\frac 1 {\sqrt
2}\left(\begin{array}{cccc}
1 & 0 & 0 & e^{-i\varphi} \\
0 & 1& \pm 1 & 0 \\
0 & \mp 1 & 1 & 0 \\
- e^{i\varphi}& 0 & 0 & 1 \end{array}\right). \en

\subsection{Solutions of the eight-vertex model for $w_3=w_4$ (II)}

Imposing $w_5=w_6$ and $w_3=w_4$ in the given eight-vertex model.
In the first case of $\frac {w_1} {w_5}=2-t$, set
$z=(t^2-2t+2)^\frac 1 2$. The BGR $b_\pm$-matrices take the form
\eq b_{\pm}=\left(\begin{array}{cccc}
2-t & 0 & 0 & q \\
0 & 1 & \pm z & 0 \\
0 & \pm z & 1 & 0 \\
q^{-1} & 0 & 0 & t
\end{array}\right).
\en It has two distinct eigenvalues $1\pm z$ and so via
Yang--Baxterization the $\check{R}(x)$-matrices are obtained by
\eq \label{eightv2} \check{R}_{\pm}(x) = \left(\begin{array}{cccc}
2-t (1- x) & 0 & 0 & q (1-x) \\
0 & 1+x & \pm z (1-x) & 0 \\
0 & \pm z (1-x) & 1+x & 0 \\
q^{-1} (1-x) & 0 & 0 &  2 x+t (1- x)
\end{array}\right).\en

For simplicity, assume $t$ real and then $z$ positive real. The
unitarity condition (\ref{unitarity}) leads to the following
equations, \eq \left\{\begin{array}{ccc}
 \|(2-t)+t x\|^2+ q\bar q\|1-x\|^2 &=& \rho \\
 \bar q^{-1} ((2-t)+t x ) (1-\bar x)+ q (1-x) (t+(2-t)\bar x) &=& 0 \\
\|1+x\|^2+z^2\,\|1-x\|^2 &=& \rho \\
z (1+x) (1-\bar x)+ z (1-x)(1+\bar x) &=& 0  \\
q^{-1}\bar q^{-1}\|1-x\|^2+\|t+(2-t) x\|^2 &=& \rho\end{array}
\right.. \en Analyzing the fourth equation leads to $\|x\|^2=1$,
while comparing the first equation and the third one gives
$\|q\|^2=1$. They survive the remaining equations. The
normalization factor $\rho$ is given by \eq \label{ucv2} \rho=2
(1+z^2)+(1-z^2)( x+\bar x)=4+(t-1)^2 (2-x-\bar x). \en

\subsection{Solutions of the eight-vertex model for $w_3=w_4$ (III)}

Following the subsection ($4.2$), consider the second case of
$w_3=w_4$, namely take $\frac {w_1} {w_5}=t$ so that $z=\pm t$.
The corresponding BGR $b$-matrix has the form \eq \label{bgr3}
b_{\pm}=\left(\begin{array}{cccc}
t & 0 & 0 & q \\
0 & 1 & \pm t & 0 \\
0 &\pm  t & 1 & 0 \\
q^{-1} & 0 & 0 & t
\end{array}\right),
\en which has three distinct eigenvalues. It needs two types of
Yang--Baxterization. The first case is essentially the same as the
case of two distinct eigenvalues, while the second case is treated
in detail in Appendix A.2.

With Yang--Baxterization of the first case, we have \eq
\label{eightv3} \check{R}_{\pm}(x)= \left(\begin{array}{cccc}
t (1- x) & 0 & 0 & q (1+x) \\
0 & 1+x & \pm t(1-x) & 0 \\
0 & \pm t (1-x) & 1+x & 0 \\
q^{-1} (1+x) & 0 & 0 &  t (1- x)
\end{array}\right).\en  The unitarity condition
(\ref{unitarity}) shows \eq  \left\{\begin{array}{ccc}
\|t\|^2\|1-x\|^2+q\bar
q\|1+x\|^2 &=& \rho \\
 \|t\|^2\|1-x\|^2+\|1+x\|^2 &=& \rho  \\
 \|t\|^2\|1-x\|^2+q^{-1}\bar q^{-1}\|1+x\|^2 &=& \rho  \\
 t(1-x)(1+\bar x)+\bar t (1+x)(1-\bar x) &=& 0 \\
 \bar q^{-1} t (1-x)(1+\bar x)+ q \bar t (1+x) (1-\bar x ) &=& 0\end{array}\right..
\en  The first three equations give $\|q\|^2=1$. Simplifying the
fourth equation gives
 \eq \|x\|^2=1+\frac {1-t_1^2} {1+t_1^2} (x-\bar x) \en
where $t_1^2=\frac {t^2}{\|t\|^2}$.

Setting $t=\|t\|e^{i\phi}$ leads to  \eq
\|x\|^2=1-i\,\tan{\phi}\,\, (x-\bar x) \en and setting $x=a+i\,b$
gives us \eq a^2+(b-\tan{\phi})^2=\sec^2{\phi} \en so that the
normalization factor $\rho$ has the form \eq \label{ucv3}
 \rho=2(1+ \|t\|^2)-i \tan{\phi}\,\,(x-\bar x)(1+\|t\|^2)+(1-
 \|t\|^2)(x+\bar x).
\en Assume $t$ real,  namely $\phi=0$, then $\|x\|^2=1$ and the
normalization factor $\rho$ is given by \eq \rho=t^2(2-x-\bar
x)+2+x+\bar x \en which is non-negative since $\|x\|=1$.

\subsection{Solutions of the eight-vertex model for $w_3=w_4$ (IV)}

Following the subsection ($4.3$), the second case for the BGR
$b$-matrix (\ref{bgr3}) having three distinct eigenvalues gives
the following $\check{R}(x)$-matrix: \eqa
\label{eightv4}\check{R}_{\pm}(x) &=& \left(\begin{array}{cccc}
t (1+ x) g_1 & 0 & 0 & q (1-x) g_1 \\
0 & (1+x) g_2 & \pm t(1-x) g_2 & 0 \\
0 & \pm t (1-x) g_2 & (1+x) g_2 & 0 \\
q^{-1} (1-x) g_1 & 0 & 0 &  t (1+ x) g_1
\end{array}\right), \nonumber\\
&\Longleftrightarrow& \left(\begin{array}{cccc}
t (1+ x)  & 0 & 0 & q (1-x) \\
0 & (1+x) g & \pm t(1-x) g & 0 \\
0 & \pm t (1-x) g & (1+x) g & 0 \\
q^{-1} (1-x)  & 0 & 0 &  t (1+ x)
\end{array}\right),
\ena where the symbols $g_1$, $g_2$ and $g$ are respectively
defined by \eq g_1=1+t+x(1-t),\qquad  g_2=1+t-x (1-t), \qquad
 g=\frac {g_2} {g_1}. \en

 The unitarity condition (\ref{unitarity}) delivers us the following
equations \eq \left\{
\begin{array}{ccc}\|t\|^2\,\|1+x\|^2+\|q\|^2\|1-x\|^2 &=& \rho\\
t \bar q^{-1} (1+x)(1-\bar x)+\bar t q (1-x)(1+\bar x) &=&0\\
q^{-1} \bar q^{-1} \|1-x\|^2+\|t\|^2 \|1+x\|^2 &=&  \rho\\
 \|g\|^2 \|1+x\|^2+\|t\|^2 \|g\|^2 \|1-x\|^2 &=& \rho\\
\bar t \|g\|^2 (1+x) (1-\bar x )+ t \|g\|^2 (1+\bar x) (1- x) &=&
0 \end{array}\right.. \en Identifying the first equation with the
third one requires $\|q\|^2=1$. Identifying the second equation
with the last one shows \eq (t-\bar t )(x-\bar x)=0. \en Focus on
the second equation: as $t$ is real, we have $\|x\|=1$; as $x$ is
real, we have $t=-\bar t$.

Comparing the first equation with the fourth one leads to the
normalization factor $\rho$. For real $t$, it takes \eq
\label{ucv41} \rho=\|g_2\|^2=2(1+t^2)-(1-t^2)(x+\bar x);\en for
pure imaginary $t$, it gives \eq \label{ucv42} \rho=
\|g_2\|^2=(1-x)^2+\|t\|^2 (1+x)^2. \en

\subsection{Two types of unitarity conditions}

In order to obtain the time-evolution of quantum entangled state
determined by the unitary $\check{R}(x)$-matrix, it is better to
consider the ordinary unitarity condition (\ref{unitarity})
instead of the unitarity condition
 $\check R(x)\check{R}(x^{-1})\propto 1\!\! 1$.

For the first type of the eight-vertex model (\ref{belltype1}), we
have
 \eq \check R(x)\check{R}(x^{-1})=2\,( x+ x^{-1}) 1\!\!1, \en
 which is incompatible with the normalization factor $\rho$ by
 \eq
 \rho=R(x)\check{R}^\dag(\bar x)=2(1+x^2) 1\!\!1 \en except $x=1$. For
the second type of the eight-vertex model (\ref{eightv2}), we have
\eq \check R(x)\check{R}(x^{-1})=2\,(1+z^2)+(1-z^2)(x+x^{-1}) \en
which is compatible with the normalization factor (\ref{ucv2}) for
real $t$. For the third type of the eight-vertex model
(\ref{eightv3}), we have \eq \check
R(x)\check{R}(x^{-1})=2(1+t^2)+(1-t^2)(x+x^{-1}) \en which is
obviously an special example of the normalization factor
(\ref{ucv3}). For the fourth type of the eight-vertex model
(\ref{eightv4}), we have
 \eq \check
R(x)\check{R}(x^{-1})=2(1+t^2)+(t^2-1)(x+x^{-1}) \en  which is the
same as the normalization factor ({\ref{ucv41}) but not
({\ref{ucv42}).

 \section{The $\check{R}(x)$-matrices as universal quantum gates}

In this section, we will view unitary $\check{R}(x)$-matrices as
universal quantum gates with the help of the Brylinski's theorem
\cite{BB}. It is a natural generalization of the argument regarding
a unitary braiding operator as a universal quantum gate. Therefore
quantum entanglements not only see topological entanglements or
topological invariants but also know geometric information or
geometric invariants hidden in unitary solutions of the QYBE
(\ref{qybe}).

A pure state $|\psi\rangle$ denoted by
 \eq
|\psi\rangle=\sum^{1}_{i,j=0}a_{ij}|ij\rangle, \qquad
|ij\rangle=|i\rangle \otimes |j\rangle
 \en
is entangled if $a_{00} a_{11}\neq a_{01} a_{10}$. The entanglement
of $| \psi \rangle$ is equivalent to the statement that $| \psi
\rangle$ is not the tensor product of two one-qubit states
\cite{werner}.

 The $\check{R}$-matrix having the form
 \eq
\check{R}=\left(\begin{array}{cccc} \check{R}^{00}_{00} &
\check{R}^{00}_{01} & \check{R}^{00}_{10}&
\check{R}^{00}_{11} \\[2mm]
\check{R}^{01}_{00} & \check{R}^{01}_{01} & \check{R}^{01}_{10}&
\check{R}^{01}_{11}
\\[2mm]
\check{R}^{10}_{00} & \check{R}^{10}_{01} & \check{R}^{10}_{10}&
\check{R}^{10}_{11}
\\[2mm]
\check{R}^{11}_{00} & \check{R}^{11}_{01} & \check{R}^{11}_{10}&
\check{R}^{11}_{11}
\end{array}\right),
 \en
acts on the tensor product $|i\rangle\otimes |j\rangle$ via the
formula \eq \check{R}|ij\rangle=\sum_{k=0}^{1} \sum_{l=0}^{1}
\check{R}^{kl}_{ij}\,|kl\rangle \en where $i,j,k,l$ take either $0$
or $1$.  The Brylinski's theorem \cite{BB} says that it is a
universal quantum gate when it is a quantum entangling operator
which transforms the tensor product $|\psi_{tp}\rangle$ into an
entangling state $\check{R}|\psi_{tp}\rangle$ given by \eq
 \check{R} |\psi_{pt}\rangle =\sum^{1}_{i,j=0}\sum_{k,l=0}^{1}
 \check{R}^{kl}_{ij} a_{ij} |kl\rangle= \sum^{1}_{k,l=0}\,b_{kl}|kl\rangle
\en where the coefficients $a_{ij}$ satisfy $a_{00} a_{11}=a_{01}
a_{10}$ and the coefficients $b_{kl}$ are defined by \eq
b_{kl}=\sum_{i,j=0}^{1}
 \check{R}^{kl}_{ij} a_{ij}  \en
 satisfying $b_{00} b_{11}\neq b_{01} b_{10}$.

Introduce the four dimensional vectors $\vec a$, $\vec b$ and
$\vec r_{ij}$ as \eq \vec a=\left(\begin{array}{c} a_{00}\\
a_{01}\\ a_{10} \\ a_{11}
\end{array}\right), \qquad \vec b=\left(\begin{array}{c} b_{00}\\
b_{01}\\ b_{10} \\ b_{11}
\end{array}\right),\qquad \vec r^T_{ij}=\left(\begin{array}{c} \check{R}^{ij}_{00}\\
\check{R}^{ij}_{01}\\ \check{R}^{ij}_{10} \\ \check{R}^{ij}_{11}
\end{array}\right)
\en where the upper index $T$ denotes the transpose of the vector
$\vec r_{ij}$, so that \eq \check R\,\, \vec a=\vec b, \qquad
b_{ij}= \vec r_{ij}\cdot \vec a. \en For convenience of the
following discussion, consider the $2\times 2$ matrix $A$ instead
of the vector $\vec a$ and the $2\times 2$ matrix $B$ instead of
the vector $\vec b$ as follows \eq A=\left(\begin{array}{cc}
a_{00}& a_{01}\\ a_{10} & a_{11}\end{array} \right), \qquad
B=\left(\begin{array}{cc} b_{00}& b_{01}\\ b_{10} & b_{11}
\end{array}\right) \en

The criteria of quantum entanglement for the quantum state
determined by the matrix $A$ ($B$) is that the determinant of the
matrix $A$ ($B$) is not zero, namely, \eq Det(A)=a_{00}
a_{11}-a_{01} a_{10}\neq 0, \qquad  Det(B)=b_{00} b_{11}-b_{01}
b_{10}\neq 0
 \en
where the determinant $Det(A)$ (or $Det(B)$) can be called as the
concurrence of the corresponding quantum state, see \cite{wootters1,
wootters2}. Here in order to judge whether the unitary
$\check{R}(x)$-matrix is a universal quantum gate with the
Brylinski's theorem \cite{BB}, we choose $Det(A)=0$ (so that the
initial state is unentangled).

\subsection{The case for solutions of the six-vertex models}

For the non-standard representation (\ref{rrmatrixone}) of the
six-vertex model, the vector $\vec{b}$ is obtained to be
\eq \left(\begin{array}{c} b_{00} \\ b_{01} \\
 b_{10} \\ b_{11} \end{array}\right)=
\left(\begin{array}{l}
 \sinh(\gamma-i\theta)\,\, a_{00}
\\  e^{i\theta}\sinh{\gamma}\,\, a_{01}- i \sin\theta\,\, a_{10}
 \\  -i\sin\theta\,\, a_{01} + e^{-i\theta}\sinh\gamma\,\, a_{10}
\\  \sinh(\gamma+i\theta)\,\, a_{11}
\end{array}\right)
\en which gives \eqa
 b_{00} b_{11} &=& (\sinh^2\gamma+\sin^2\theta)\, a_{00} a_{11}, \nonumber\\
 b_{01} b_{10} &=&(\sinh^2\gamma-\sin^2\theta)\, a_{01} a_{10}-
 i \sin\theta\sinh\gamma (a_{01}^2  e^{i\theta}+ a_{10}^2
 e^{-i\theta}).  \ena So the criteria of quantum entanglement has the form
 \eq Det(B)=\sin\theta [ 2 a_{00} a_{11} \sin\theta\,+\,i(a^2_{01}
 e^{i\theta}+ a_{10}^2 e^{-i\theta}) \sinh\gamma ]\neq 0.\en

Consider the case of the spectral parameter $x\neq 1$, namely
$\sin\theta\neq 0$. The choice: $a_{00}=a_{10}=0$, $a_{01}\neq 0$
and $\gamma \neq 0$ satisfies $Det(B)\neq 0$. When $\gamma=0$,
namely the deformation parameter $q=1$, the criteria of quantum
entanglement requires $a_{00} a_{11}\neq 0$. Hence the unitary
$\check{R}(\theta)$-matrix for the non-standard representation of
the six-vertex model is a universal quantum gate except for $x=1$.

For the standard representation (\ref{rmatrixtwo}) of the
six-vertex model, the determinant of the matrix $B$ is a
difference between two terms $b_{00} b_{11}$ and $b_{01} b_{10}$
given by \eqa
 b_{00} b_{11} &=& \sinh(\gamma-i\theta)\sinh(\gamma-i\theta)\,a_{00} a_{11},
 \nonumber\\ b_{01} b_{10} &=&(\sinh^2\gamma-\sin^2\theta)\, a_{01} a_{10}-
 i \sin\theta\sinh\gamma (a_{01}^2  e^{i\theta}+ a_{10}^2
 e^{-i\theta})\ena The
choice:  $a_{00}=a_{10}=0$, $a_{01}\neq 0$ and $\gamma \neq 0$
satisfies $Det(B)\neq 0$. But the case of $\gamma=0$ leads to
$Det(B)=0$. Hence the unitary $\check{R}(\theta)$-matrix for the
standard representation of the six-vertex model is a universal
quantum gate except for $x=1$ or $q=1$.

To distinguish the non-standard representation from the standard
representation in a clear way, the $\check{R}(x)$-matrices of
$q=1$ are given respectively by \eq
\check{R}_{non}=\left(\begin{array}{cccc}
1 & 0 & 0 & 0 \\
0 & 0 & 1 & 0 \\
0 & 1 & 0 & 0 \\
0 & 0 & 0 & -1
\end{array}\right), \qquad \check{R}_{standard}=\left(\begin{array}{cccc}
1 & 0 & 0 & 0 \\
0 & 0 & 1 & 0 \\
0 & 1 & 0 & 0 \\
0 & 0 & 0 & 1
\end{array}\right)
\en in which the non-standard one for constructing the Alexander
polynomial is a universal quantum gate and the standard one for
constructing the Jones polynomial is not a universal quantum gate
\cite{kauffman0}.

\subsection{The case for solutions of the eight-vertex model (I)}

For simplicity of analyzing the criteria of quantum entanglement,
we introduce the new variable $u, v$ instead of the spectral
parameter $x, y$ so that \eq u=\frac {1-x} {1+x},\qquad v=\frac
{1-y} {1+y},\qquad \frac {1-x y} {1+x y}=\frac {u+v} {1+u v}\en
which suggest the following Yang--Baxter-like equation \eq
 \check{R}_{12}(u)\,\check{R}_{23}(\frac {u+v} {1+u v})\,
 \check{R}_{12}(v)= \check{R}_{23}(u)\,\check{R}_{12}( \frac
{u+v} {1+u v})\,\check{R}_{23}(v). \en Specifying
$x=e^{i\theta_1}$ and $y=e^{i\theta_2}$, the parameters in the
above equation have the forms\eq u=-i\tan\theta_1,\,\,
v=-i\tan\theta_2,\, \,\frac {u+v} {1+u
v}=-i\tan(\theta_1+\theta_2).
 \en

The unitary $\check{R}(u)$-matrix being a universal quantum gate
suggests that the unitary $\check{R}(x)$-matrix is a universal
quantum gate since the determinant $Det(B)$ are not vanishing in
both cases. In terms of the new variable $u$, the unitary
$\check{R}_\pm(u)$-matrix for the first type of solution
(\ref{belltype1}) of the eight-vertex model has the form \eq
\check{R}_\pm(u)
=\left(\begin{array}{cccc} 1 & 0 & 0 & q u \\
 0 & 1 & \pm  u & 0 \\
 0 &  \mp  u & 1 & 0 \\
 -q^{-1} u & 0 & 0 & 1
\end{array}\right).
\en The corresponding matrix $B^\pm$ is given by \eq
\left(\begin{array}{cc} b^\pm_{00} & b^{\pm}_{01} \\ b^{\pm}_{10}
& b^\pm_{11}
\end{array}\right)= \left(\begin{array}{cc}
a_{00}+q u\,\, a_{11}  &  a_{01}\pm u\,\, a_{10}
 \\  \mp u\,\, a_{01} +a_{10}
& -q^{-1} u \,\,a_{00} + a_{11}
\end{array}\right).
\en It has the determinant  \eq
 Det(B^\pm)=u (q\, a_{11}^2-q^{-1}\, a_{00}^2 \pm a_{01}^2 \mp a_{10}^2)\en
which is not zero for the case of
 $u\neq 0$ ($x\neq 1$), $a_{00}=a_{01}=0$
 and $a_{10}^2\neq q a_{11}^2 $. Hence the unitary
 $\check{R}_\pm(u)$-matrix
 is a universal quantum gate except $u=0$, that is to say
the unitary $\check{R}_\pm(x)$-matrix is a universal quantum gate
except $x=1$.

 In terms of the coefficients $a, b, c, d$ given by
 \eq \label{coefficients} a_{00}=ac, \,\, a_{01}=ad,\,\,
 a_{10}=bc,\,\, a_{11}=bd, \en instead of $a_{ij}$, the criteria of
 quantum entanglement leads to
 \eq (d^2\mp q^{-1} c^2)(q b^2\pm a^2)\neq 0. \en
 Consider which type of quantum states violate the criteria of
 quantum entanglements. For the $\check{R}_+$-matrix, we have
 \eq d^2 = q^{-1} c^2,\qquad \textrm{or}\qquad a^2=-q\, b^2, \en
 while for the $\check{R}_-$-matrix, we have
 \eq d^2=- q^{-1} c^2,\qquad  \textrm{or} \qquad a^2=q\,b^2. \en
 Therefore, although a universal quantum gate is identified as
 an entangling operator, it is also able to transform a specified
unentangled state (a tensor product of quantum states) into
another unentangled state.

\subsection{The case for solutions of the eight-vertex model (II)}

In terms of the new variable $u$, the corresponding
$\check{R}_\pm(u)$-matrix has the form \eq \check{R}_\pm(u)
=\left(\begin{array}{cccc} 1+(1-t) u & 0 & 0 & q u \\
 0 & 1 & \pm z u & 0 \\
 0 &  \pm z u & 1 & 0 \\
 q^{-1} u & 0 & 0 & 1+(t-1) u
\end{array}\right)
\en  so that the vector $\vec b^\pm$ are given by \eq
\left(\begin{array}{c} b^\pm_{00} \\ b^{\pm}_{01} \\ b^{\pm}_{10}
\\ b^\pm_{11}
\end{array}\right)= \left(\begin{array}{c}
(1+(1-t) u) a_{00}+q u\,\, a_{11} \\  a_{01}\,\pm z u\, a_{10}
 \\  \pm z u\,\, a_{01} +a_{10}
\\ q^{-1} u \,\,a_{00} + (1+(t-1) u) a_{11}
\end{array}\right)
\en which leads to \eqa b^\pm_{00} b^\pm_{11} &=& (1+(2-z^2)
u^2)\, a_{00}
a_{11} \nonumber\\
& &+u(q^{-1} (1+(1-t)u)a^2_{00}  +q (1+(t-1)u) a^2_{11}), \nonumber\\
 b^\pm_{01} b^\pm_{10} &=& (1+z^2 u^2)\, a_{01} a_{10} \pm u z (a^2_{01}+a^2_{10}).
\ena To satisfy $Det(B)\neq 0$, choose $u\neq 0$,
$a_{00}=a_{01}=a_{11}=0$ but $a_{10}\neq 0$. Hence the unitary
$\check{R}_\pm(x)$-matrix (\ref{eightv2}) is a universal quantum
gate for real $t$ and $x\neq 1$.

\subsection{The case for solutions of the eight-vertex model (III)}

With the new variable $u$, the $\check{R}_\pm(u)$-matrix is given
by \eq \check{R}_\pm(u)
=\left(\begin{array}{cccc} t u & 0 & 0 & q  \\
 0 & 1 & \pm t u & 0 \\
 0 &  \pm t u & 1 & 0 \\
 q^{-1}  & 0 & 0 & t u
\end{array}\right)\en
which says \eqa
 b^\pm_{00} b^\pm_{11} &=& (1+t^2 u^2)\, a_{00} a_{11}+ u t (q a^2_{11}+ q^{-1}
 a^2_{00}),
 \nonumber\\
 b^\pm_{01} b^\pm_{10} &=& (1+t^2 u^2)\, a_{00} a_{11} \pm u t
 (a^2_{01}+a^2_{10}).
 \ena
So the unitary $\check{R}(x)$-matrix (\ref{eightv3}) is a
universal quantum gate except $tu=0$.

The criteria of quantum entanglement in terms of the coefficients
(\ref{coefficients}) has the form \eq (a^2\mp b^2 q^{-1}) (c^2 q
\mp d^2 ) \neq 0. \en For the unitary $\check{R}_\pm(x)$-matrix,
the quantum state specified by \eq a^2=\pm b^2 q^{-1},\qquad
\textrm{or} \qquad d^2=\pm c^2 q \en is not entangling even under
the action of the unitary $\check{R}_\pm(x)$-matrix.

\subsection{The case for solutions of the eight-vertex model (IV)}

With the new variable $u$, the $\check{R}_\pm(u)$-matrix has the
form \eq \check{R}_\pm(u)
=\left(\begin{array}{cccc} t (1+t u) & 0 & 0 & q u (1+t u)  \\
 0 & u+t & \pm t u (u+t ) & 0 \\
 0 &  \pm t u (u+t ) & u+t & 0 \\
 q^{-1} u(1+t u)  & 0 & 0 & t (1+t u)
\end{array}\right)\en
which gives us \eqa
 b^\pm_{00} b^\pm_{11} &=& (t^2+u^2)(1+t u)^2\, a_{00} a_{11}
 + u t (1+t u)^2 (q^{-1} a^2_{00}+q a^2_{11}),
 \nonumber\\
 b^\pm_{01} b^\pm_{10} &=& (1+t^2 u^2)(u+t)^2\,a_{00} a_{11}
 \pm u t (u+t)^2 (a^2_{01}+
 a^2_{10}).
 \ena
It can be observed that the unitary $\check{R}_\pm(x)$-matrix
(\ref{eightv4}) is a universal quantum gate for $ut\neq 0$.

\section{The constructions of the Hamiltonian}

In this section, we present a method of constructing the
Hamiltonian from the unitary $\check{R}(x)$-matrix
($\check{R}(\theta)$-matrix) for the six-vertex and eight-vertex
model. The comments on our construction are given in the last
subsection.

The wave function $\psi(x)$ is specified by the unitary
$\check{R}(x)$-matrix, with $\psi(x)=\check{R}(x)\psi$, the pure
state $\psi$ independent of the time (or the spectral parameter
$x$). Hence we obtain the Shr{\" o}dinger equation corresponding
to the time evolution of $\psi(x)$ controlled by the unitary
$\check{R}(x)$-matrix, \eq i\,\frac {\partial \psi(x)} {\partial
x}=H(x)\psi(x), \qquad H(x)=i\,\frac {\partial \check{R}(x)}
{\partial{x}} \check{R}^{-1}(x). \en Here the unitary
$\check{R}(x)$-matrix has to be the form containing the
normalization factor, namely $\rho^{-\frac 1 2}\check{R}(x)$
satisfying $(\rho^{-\frac 1 2}\check{R})^{-1}=(\rho^{-\frac 1
2}\check{R})^\dag$. The ``time-dependent" Hamiltonian $H(x)$ is
obtained to be \eq H(x)=i\,\frac {\partial (\rho^{-\frac 1
2}\check{R})}{\partial x}(\rho^{-\frac 1
2}\check{R})^{-1}(x)=i\,\frac {\partial(\rho^{-\frac 1
2}\check{R}) }{\partial x} \rho^{-\frac 1 2}\check{R}^{\dag}(x).
\en It is expanded into the form \eqa H(x) &=&  i ({\partial_x
\ln\rho^{-\frac 1 2}(x)} + {\partial_x \check{R}(x)}
\check{R}^{-1}(x))
 \nonumber\\
&=& i\rho^{-1}(x)(-\frac 1 2\frac {\partial\rho(x)} {\partial
x}+\frac {\partial\check{R}(x)}{\partial x} \check{R}^{\dag}(x)
)\ena where the formula in terms of $\check{R}^{\dag}(x)$ is
applied in the following since the calculation of
$\check{R}^{\dag}(x)$ is easier than that of $\check{R}^{-1}(x)$.

In this paper, the spectral parameter $x$ often takes $\|x\|=1$,
namely $x=e^{i\theta}$ so the Hamiltonian $H(\theta)$ has another
form \eq H(\theta)=i\rho^{-1}(\theta)(-\frac 1 2\frac
{\partial\rho(\theta)} {\partial \theta}+\frac
{\partial\check{R}(\theta)}{\partial \theta}
\check{R}^{\dag}(\theta) ) \en which leads to
 the following Schr{\"o}dinger equation \eq i\,\frac {\partial \psi(\theta)}
{\partial \theta}=H(\theta) \psi(\theta).\en

\subsection{The case for solutions of the six-vertex models}

For the non-standard representation (\ref{rrmatrixone}) of the
six-vertex model, the Hamiltonian $H(\theta)$ is constructed as
follows \eq \label{hamil1} H(\theta)
=\frac {\sinh\gamma} \rho\left(\begin{array}{cccc} \coth\gamma & 0 & 0 & 0  \\
 0 & -\sinh\gamma & 1 & 0 \\
 0 &  1 & \sinh\gamma & 0 \\
0  & 0 & 0 & -\coth\gamma
\end{array}\right)\en
where the following formulas are used in calculation \eqa
\sinh(a+b)&=&\sinh a\,\coth b+ \sinh b\,\coth a, \qquad \sinh
ia=i\sin a, \nonumber\\
\coth(a+b) &=& \coth a \coth b+\sinh a\sinh b, \qquad \coth
ia=\cos a. \ena

Here the Pauli matrices $\sigma_x$, $\sigma_y$ and $\sigma_z$ are
set up as usual \eq \sigma_x=\left(\begin{array}{cc} 0 & 1\\
1 & 0 \end{array} \right), \qquad
\sigma_y=\left(\begin{array}{cc} 0 & -i\\
i & 0 \end{array} \right), \qquad
\sigma_z=\left(\begin{array}{cc} 1 & 0\\
0 & -1 \end{array} \right).\en In addition, the new matrices
$\sigma_\pm$ are introduced by $\sigma_\pm=\frac 1 2(\sigma_x \pm
i \sigma_y)$: \eq
\sigma_+=\left(\begin{array}{cc} 0 & 1\\
0 & 0 \end{array} \right), \qquad
\sigma_-=\left(\begin{array}{cc} 0 & 0\\
0 & 1 \end{array} \right).\en To represent the Hamiltonian
(\ref{hamil1}) in terms of the Pauli matrices, the formulas like
\eq \frac 1 2 (1\!\! 1\otimes \sigma_z+\sigma_z\otimes 1\!\!
1)=\left(\begin{array}{cccc} 1 & 0 & 0 & 0\\ 0 & 0 & 0 & 0 \\ 0 &
0 & 0& 0 \\ 0 & 0 & 0 & -1
\end{array}\right), \,\,
\sigma_+ \otimes \sigma_-=\left(\begin{array}{cccc} 0 & 0 & 0 & 0\\
0 & 0 & 1 & 0 \\ 0 & 0 & 0& 0 \\ 0 & 0 & 0 & 0
\end{array}\right)
 \en are often used.

The Hamiltonian (\ref{hamil1}) has another form by the Pauli
matrices \eqa \label{hamil11} H(\theta)&=& \frac {\sinh\gamma}
{2\rho} [\coth\gamma(1\!\! 1\otimes \sigma_z+\sigma_z\otimes 1\!\!
1)+ \sinh\gamma(1\!\! 1\otimes
\sigma_z-\sigma_z\otimes 1\!\! 1) \nonumber\\
 & & +
(\sigma_+\otimes\sigma_-+\sigma_-\otimes\sigma_+)] \nonumber\\
&=& \frac {\sinh\gamma} {2\rho} [\coth\gamma(1\!\! 1\otimes
\sigma_z+\sigma_z\otimes 1\!\! 1)+
(\sigma_x\otimes\sigma_x+\sigma_y\otimes\sigma_y)
\nonumber\\
 & & + \sinh\gamma(1\!\! 1\otimes \sigma_z-\sigma_z\otimes 1\!\! 1)]
 \ena
where only the normalization factor $\rho$ depends on the time
variable $\theta$ by \eq \rho=\sin^2\theta+\sinh^2\gamma. \en

For the standard representation (\ref{rmatrixtwo}) of the
six-vertex model, the Hamiltonian $H(\theta)$ is obtained to be
\eq H(\theta)=
\frac {\sinh\gamma} \rho\left(\begin{array}{cccc} \coth\gamma & 0 & 0 & 0  \\
 0 & -\sinh\gamma & 1 & 0 \\
 0 &  1 & \sinh\gamma & 0 \\
0  & 0 & 0 & \coth\gamma
\end{array}\right)\en

In terms of the Pauli-matrices, it has the form \eqa H(\theta)&=&
\frac {\sinh\gamma} {2\rho} [\coth\gamma (1\!\! 1+
\sigma_z\otimes\sigma_z)+ \sinh\gamma(1\!\! 1\otimes
\sigma_z-\sigma_z\otimes 1\!\! 1) \nonumber\\
 & & +
(\sigma_+\otimes\sigma_-+\sigma_-\otimes\sigma_+)] \nonumber\\
&=& \frac {\sinh\gamma} {2\rho} [\coth\gamma 1\!\! 1 +
\sinh\gamma(1\!\! 1\otimes
\sigma_z-\sigma_z\otimes 1\!\! 1) \nonumber\\
 & & +(\sigma_x\otimes\sigma_x+\sigma_y\otimes\sigma_y
 +\coth\gamma\,\,\sigma_z\otimes\sigma_z )]
 \ena
which is different from the Hamiltonian (\ref{hamil11}) for the
non-standard representation of the six-vertex model.

\subsection{The case for solutions of the eight-vertex model (I)}

 With the unitary solution $\check{R}(x)$ (\ref{belltype1}) of the QYBE (\ref{qybe}),
 we construct the time-independent Hamiltonian $H_\pm$ having the form
  \eq H_\pm=i \frac {\partial}{\partial x}(\rho^{-\frac 1 2}\check{R}_\pm)|_{x=1}
  =-\frac i 2 b^2_\pm = \frac i 2 \left(\begin{array}{cccc}
  0 & 0 & 0 & -\,e^{-i\varphi}  \\
  0 & 0 & \mp 1 & 0     \\
  0 & \pm 1 & 0 & 0    \\
 \,e^{i\varphi}& 0 & 0 & 0 \end{array}\right). \en
Interestingly, we have the ``time-dependent" Hamiltonian
$H_\pm(x)$ by \eq H_\pm(x)=i\,\frac {\partial (\rho^{-\frac 1
2}\check{R}_\pm)  }{\partial x} \rho^{-\frac 1
2}\check{R}_\pm^{\dag}(x)=-\frac i {1+x^2}\,\, b^2_\pm \en which
derives the above Hamiltonian $H_\pm$ at $x=1$. When $x$ is real,
the Hamiltonian $H_\pm(x)$ is a Hermitian operator.

For simplicity, we study the unitary $\check{R}(\theta)$-matrix
(\ref{bellrmatrix}) to decide the unitary evolution of quantum
states. After some algebra, the Hamiltonian $H_\pm$ is obtained
 to be  \eq \label{hamiltonian}
 H_\pm(\theta)=i\rho^{-1}(\theta)(-\frac 1 2\frac
{\partial\rho(\theta)} {\partial \theta}+\frac
{\partial\check{R}_\pm(\theta)}{\partial \theta}
\check{R}_\pm^{\dag}(\theta) )=\frac i 2 \frac {\partial
x}{\partial \theta}
 H_\pm(x)=H_\pm \en
 which is independent of the time variable $\theta$.

In terms of the Pauli matrices $\sigma_x$, $\sigma_y$ and
$\sigma_z$ and $\sigma_\pm$, the Hamiltonian (\ref{hamiltonian})
has the form \eq \label{hamiltonian1} H_\pm=\frac i
2\,(-e^{-i\varphi}\sigma_+ \otimes \sigma_+ + e^{i \varphi}
\sigma_-\otimes \sigma_-\mp \sigma_+\otimes \sigma_-\pm
\sigma_-\otimes \sigma_+). \en

Introducing the two-dimensional vector $\vec \sigma$ and two unit
directional vector $\vec n_1$ and $\vec n_2$ in $xy$-plane: \eq
\vec \sigma=(\sigma_x,\sigma_y); \qquad \vec n_1=(\cos\frac
{\pi+\varphi} 2, \sin\frac {\pi+\varphi} 2),\,\, \vec
n_2=(\cos\frac \varphi 2, \sin\frac \varphi 2),  \en the
projections of the vector $\vec \sigma$ into $\vec n_1$ and $\vec
n_2$ are given by \eqa \sigma_{n_1} &=& \vec \sigma \cdot \vec
n_1=\sigma_+ e^{-\frac i 2 (\varphi+\pi)}+\sigma_- e^{\frac i 2
(\varphi+\pi)}, \nonumber\\
\sigma_{n_2} &=& \vec \sigma \cdot \vec n_2=\sigma_+ e^{-\frac i 2
\varphi}+\sigma_- e^{\frac i 2 \varphi}. \ena The Hamiltonian
(\ref{hamiltonian1}) can be recast to \eq \label{hamiltonian2}
H_+=\frac 1 2 \sigma_{n_1}\otimes \sigma_{n_2}, \qquad H_-=\frac 1
2\sigma_{n_2}\otimes \sigma_{n_1}. \en

Consider the time-evolution operator $U_\pm(\theta)$ determined by
the Hamiltonian $H_\pm$, for example, $U_+(\theta)$ given by \eq
\label{evolution1}   U_+(\theta)=e^{-\frac i 2(\sigma_{n_1}\otimes
\sigma_{n_2}) \theta } =\cos\frac \theta 2-\,i\,\sin\frac \theta
2\,\sigma_{n_1}\otimes \sigma_{n_2}. \en

\subsection{The case for solutions of the eight-vertex model (II)}

The unitary $\check{R}_\pm(x)$-matrix (\ref{eightv2}) requires
$x=e^{i\theta}$, $\|q\|^2=1$ and real $t$. The normalization
factor $\rho$ and its derivative have the forms \eq
 \rho=4+4 (t-1)^2 \sin^2\frac {\theta} 2, \qquad
 \frac {\partial \rho} {\partial \theta}=2(t-1)^2 \sin\theta. \en
After some calculation, the Hamiltonian $H_\pm(\theta)$ is
obtained to be
 \eqa H_\pm(\theta) &=&
i\rho^{-1}(\theta)(-\frac 1 2\frac {\partial\rho(\theta)}
{\partial \theta}+\frac {\partial\check{R}_\pm(\theta)}{\partial
\theta}\check{R}_\pm^{\dag}(\theta) ) \nonumber\\
 &=&-\frac {1\!\! 1} 2+ 2\rho^{-1} \left(\begin{array}{cccc}  1-t & 0 & 0 & q  \\
 0 & 0 & \pm  z & 0 \\
 0 &  \pm  z & 0 & 0 \\
 q^{-1}  & 0 & 0 & t-1
\end{array}\right).\ena
In terms of the Pauli matrices, it is shown up as follows \eqa
H_\pm(\theta) &=& -\frac 1 2 1\!\! 1+ (1-t) \rho^{-1} (1\!\!
1\otimes \sigma_z + \sigma_z\otimes 1\!\! 1) \nonumber\\
 & & +\frac 2 \rho  [q\sigma_+\otimes \sigma_+ + q^{-1} \sigma_-
 \otimes \sigma_-\pm z(\sigma_+\otimes
 \sigma_-+\sigma_-\otimes \sigma_+)].
\ena

Consider two special cases of making the normalization factor
$\rho$ independent of the time variable $\theta$. Take $\theta=0$,
namely, choose the Hamiltonian $H_\pm(\theta)$ defined as \eq
H_\pm(\theta)=i\frac {\partial}{\partial \theta}(\rho^{-\frac 1
2}\check{R}_\pm(\theta))|_{\theta=0} \en which leads to the
time-independent Hamiltonian \eqa H_\pm|_{\theta=0} &=&
 \frac 1 2  [q\sigma_+\otimes \sigma_+ + q^{-1} \sigma_-
 \otimes \sigma_-\pm z(\sigma_+\otimes
 \sigma_-+\sigma_-\otimes \sigma_+)] \nonumber\\
& & -\frac 1 2 1\!\! 1+ \frac {1-t} 4 (1\!\! 1\otimes \sigma_z +
\sigma_z\otimes 1\!\! 1). \ena Take the parameter $t=1$, the
time-independent Hamiltonian $H_\pm$ is given by \eq \label{v2h1}
H_\pm=
 \frac 1 2  [-1\!\! 1+q\sigma_+\otimes \sigma_+ + q^{-1} \sigma_-
 \otimes \sigma_- \pm ( \sigma_+\otimes
 \sigma_-+\sigma_-\otimes \sigma_+)  ].
\en

With the two unit directional vector $\vec n_1$ and $\vec n_2$ in
$xy$-plane: \eq  \vec n_1=(\cos\frac \varphi 2, \sin\frac \varphi
2),\,\, \vec n_2=(\cos\frac {\pi+\varphi} 2, \sin\frac
{\pi+\varphi} 2),\,\, q=e^{-i\varphi}  \en  the Hamiltonians
(\ref{v2h1}) have the following forms \eq
 H_+=\frac 1 2(-1\!\! 1+\sigma_{n_1}\otimes
\sigma_{n_1}), \qquad H_-=-\frac 1 2(1\!\! 1+\sigma_{n_2}\otimes
\sigma_{n_2}). \en Consider the unitary time-evolution operator
$U_\pm(\theta)$, for example, $U_+(\theta)$ given by \eq
U_+(\theta)=e^{-i H_+ \theta } =e^{\frac {i\theta} 2 }\,(\cos\frac
\theta 2-\,i\,\sin\frac \theta 2\,\sigma_{n_1}\otimes \sigma_{n_1}).
\en

\subsection{The case for solutions of the eight-vertex model (III)}

Choose real $t$, $x=e^{i\theta}$ and $\|q\|=1$ for the unitary
$\check{R}_\pm(x)$-matrix (\ref{eightv3}). The normalization factor
$\rho$ and its derivative are given by \eq \rho=4+4(t^2-1)\sin^2
{\frac \theta 2},\qquad
 \frac {\partial \rho} {\partial \theta}=2(t^2-1)\sin\theta.
 \en
The corresponding Hamiltonian $H_\pm(\theta)$ is found to be \eq
H_\pm(\theta) =-\frac 1 2 1\!\! 1+\frac {2 t} {\rho}
\left(\begin{array}{cccc}
 0 & 0 & 0 & q \\
 0 & 0 & \pm 1 & 0 \\
 0 &  \pm 1 & 0 & 0 \\
 q^{-1}  & 0 & 0 & 0
\end{array}\right)\en
which leads to the Hamiltonian represented by the Pauli matrices
\eq H_\pm(\theta)= -\frac 1 2 1\!\! 1+\frac {2t} \rho
[q\sigma_+\otimes \sigma_+ + q^{-1} \sigma_-
 \otimes \sigma_- \pm \,( \sigma_+\otimes
 \sigma_-+\sigma_-\otimes \sigma_+) ].
\en

In the case of $\theta=0$, the time-independent Hamiltonian is
given by \eq H_\pm|_{\theta=0}= -\frac 1 2 1\!\! 1+\frac t 2
[q\sigma_+\otimes \sigma_+ + q^{-1} \sigma_-
 \otimes \sigma_- \pm \,( \sigma_+\otimes
 \sigma_-+\sigma_-\otimes \sigma_+) ].
\en In the other case of $t=1$, the other time-independent
Hamiltonian $H_\pm$ has the form the same as (\ref{v2h1}).

\subsection{The case for solutions of the eight-vertex model (IV)}

Consider $x=e^{i\theta}$, $\|q\|=1$ and real $t$ for the unitary
$\check{R}_\pm(x)$-matrix (\ref{eightv4}). The factor $\|g_1\|^2$,
$\|g_2\|^2$ and the normalization factor $\rho$ take the forms
 \eqa \|g_1\|^2 &=& 2(1+t^2)+2(1-t^2)\cos\theta,
  \qquad \|g_2\|^2=2(1+t^2)-2(1-t^2)\cos\theta, \nonumber\\
 \rho &=& \|g_1\|^2\|g_2\|^2=4(1+t^2)^2-4(1-t^2)^2 \cos^2\theta.
\ena Through some calculation, the Hamiltonian $H_\pm(\theta)$ is
obtained to be\eq H_\pm(\theta) =-1\!\! 1+\frac {2 t} {\rho}
\left(\begin{array}{cccc}
 \|g_2\|^2 & 0 & 0 & q \|g_1\|^2 \\
 0 & \|g_1\|^2 & \pm \|g_2\|^2 & 0 \\
 0 &  \pm \|g_2\|^2 & \|g_1\|^2 & 0 \\
 q^{-1}\|g_1\|^2  & 0 & 0 & \|g_2\|^2
\end{array}\right)\en
which gives the following Hamiltonian as \eqa
 H_\pm(\theta) =-1\!\! 1+ \frac {2t} \rho [
 \frac 1 2 (\|g_1\|^2+\|g_2\|^2) 1\!\! 1\otimes 1\!\! 1
 + \frac 1 2 (\|g_2\|^2-\|g_1\|^2) \sigma_z\otimes \sigma_z
 \nonumber\\
 +\|g_1\|^2(q \sigma_+\otimes \sigma_+ + q^{-1} \sigma_-\otimes \sigma_-)
 \pm \|g_2\|^2 (\sigma_+\otimes \sigma_- + \sigma_-\otimes \sigma_+ )]. \ena

 The time-independent Hamiltonian obtained by taking $\theta=0$ is
 given by
  \eqa
  H_\pm|_{\theta=0} =-1\!\! 1+ \frac 1 {4t} [
  (t^2+1) 1\!\! 1\otimes 1\!\! 1
  + (t^2-1) \sigma_z\otimes \sigma_z
   \nonumber\\
 + 2 (q \sigma_+\otimes \sigma_+ + q^{-1} \sigma_-\otimes \sigma_-)
    \pm 2 t^2 (\sigma_+\otimes \sigma_- + \sigma_-\otimes \sigma_+
    )],
 \ena
 while the other time-independent Hamiltonian $H_\pm$ obtained by taking $t=1$
 is the same as (\ref{v2h1}).

 \subsection{Comments on our constructions of Hamiltonian}

The Schrodinger equation is a differential equation. It's solution
represents the evolution of the initial state (input). Here before
treating the Schrodinger equation, the time evolution of the state
(the unitary braiding operator) is known as a discrete evolution and
is determined by the unitary $\check{R}(x)$-matrix. Our problem is
to find out which type of the Schrodinger equation has evolutionary
solutions the same as the evolutions given by the unitary
$\check{R}(x)$-matrices. In our case, therefore, the Schrodinger
evolution is recognized as the unitary $\check{R}(x)$-matrix.

However, the Schrodinger equation says more than just the unitary
$\check{R}(x)$-matrix. It leads us to study the physics behind it,
and gives us new unitary solutions which are not necessarily
solutions to the QYBE. The construction of a Hamiltonian from the
unitary $\check{R}(x)$-matrix shows that the QYBE is a physical
subject. In history, the Schrodinger equation is the starting
point and how to solve the Schrodinger equation is a central
topic. Now we try to recover(relate) physics from (to) the unitary
$\check{R}(x)$-matrix.

In our case, the construction of the Hamiltonian in terms of the
unitary $\check{R}(x)$-matrix follows a traditional approach in
literature on the QYBE. There are two ways of choosing the
time-variable. Each choice is determined by a corresponding
purpose. We explain the spectral parameter $x$ ($\theta$) as the
time variable. That is to say that we choose the time-evolution of
quantum state (the unitary braiding operator) as the unitary
$\check{R}(x)$-matrix. This is a natural and necessary choice in
the sense of regarding unitary $\check{R}$-matrices as universal
quantum gates. We want to see the evolution of unitary braiding
operator or quantum state.

Indeed in the literature, the ordinary time in space-time as the
time variable, the spectral parameter can be explained as the
momentum, but the time-evolution of quantum state determined by
the Hamiltonian can not be identified with the unitary
$\check{R}(x)$-matrix. The physics in the case is well-known such
as XXX model (and so on).

Finally it seems that the Schrodinger equation in our construction
does not have space variables like the ordinary Schrodinger
equation has. We could explain that we have physics on lattices of
space: discrete physics. It seems that physics in our case is
close to that on spin chains like XXX model (and so on).

 \section{The CNOT gates via the $\check{R}$-matrix}

The gate $G$ is universal for quantum computation (or just
universal) if $G$ together with local unitary transformations
(unitary transformations from $V$ to $V$) generates all unitary
transformations of the complex vector space of dimension $2^{n}$ to
itself. It is well-known \cite{nielsen} that the CNOT gate is a
universal gate. \bigbreak

\noindent In \cite{kauffman8},  Kauffman and Lomonaco prove the
following result.

\bigbreak

\noindent {\bf Theorem 1}. Let \eq \label{theorem} \check{R} =
\left(
\begin{array}{cccc}
1/\sqrt{2} & 0 & 0  & 1/\sqrt{2}\\
0  & 1/\sqrt{2} & -1/\sqrt{2} & 0\\
0 & 1/\sqrt{2} & 1/\sqrt{2} & 0 \\
-1/\sqrt{2} & 0 & 0 & 1/\sqrt{2}\\
\end{array} \right)\en be the above unitary solution to the braid
relation (\ref{bgr}). Then $\check{R}$ is a universal gate. The
proof below (repeated from \cite{kauffman8}) gives a specific
expression for the CNOT gate in terms of  $\check{R}.$ \bigbreak

 \noindent {\bf Proof.} This result follows at once from the
Brylinksis' theorem \cite{BB}, since $\check{R}$ is highly
entangling. For a direct computational proof, it suffices to show
that the CNOT gate can be generated from $\check{R}$ and local
unitary transformations. Let \eqa \label{delta} \alpha &=& \left(
\begin{array}{cc}
1/\sqrt{2} & 1/\sqrt{2}\\
1/\sqrt{2} & -1/\sqrt{2}\\
\end{array} \right), \qquad
\beta = \left( \begin{array}{cc}
-1/\sqrt{2} & 1/\sqrt{2}\\
i/\sqrt{2} & i/\sqrt{2}\\
\end{array} \right) \nonumber\\
\gamma &=& \left( \begin{array}{cc}
1/\sqrt{2} & i/\sqrt{2}\\
1/\sqrt{2} & -i/\sqrt{2}\\
\end{array} \right),\qquad
\delta = \left( \begin{array}{cc}
1 & 0\\
0 & i\\
\end{array} \right).\ena  Let $M= \alpha \otimes \beta$ and $N= -\gamma \otimes \delta.$
Then it is straightforward to verify that
$$\textrm{CNOT} = M\cdot \check{R} \cdot N.$$ This completes the proof. $\hfill \Box $
\bigbreak

 We now show how Yang--Baxterization illuminates the structure of this
Bell basis transformation. We discuss physics related to the
 time-evolution of the universal quantum gate determined by the unitary
$\check{R}(\theta)$-matrix
 (\ref{bellrmatrix}). The braid
 group representation $b_\pm(\varphi)$-matrix (\ref{bgrphi}) yields
 the Bell states with the phase factor $e^{i\varphi}$,
\eq b_\pm(\varphi) \left(\begin{array}{c} |0 0\rangle \\| 0 1 \rangle \\
|1 0\rangle \\ |1 1\rangle
\end{array}\right)=\frac 1 {\sqrt{2}}
 \left(\begin{array}{c} |0 0\rangle-e^{i\varphi}|1 1\rangle \\
 |0 1 \rangle \mp |1 0 \rangle \\ \pm|0 1\rangle + | 1 0 \rangle\\
 e^{-i\varphi} |0 0\rangle+|11\rangle
\end{array}\right)
\en which shows that $\varphi=0$ leads to the Bell states, the
maximum of entangled states, \eq \frac 1 {\sqrt{2}} (|0 0\rangle
\pm |1 1\rangle), \qquad \frac 1 {\sqrt{2}} (| 1 0 \rangle \pm |0
1 \rangle). \en

 In terms of the Hamiltonian $H_\pm$ (\ref{hamiltonian}),
 the $\check R_\pm(\theta)$-matrix (\ref{bellrmatrix}) has the form
 \eq \check{R}_\pm(\theta)=\cos(\frac \pi 4-\theta)+ 2\,i\,
 \sin(\frac \pi 4-\theta)\,H_{\pm}=e^{i (\frac \pi
 2-2\,\theta) H_\pm} \en which can be also used to construct the
 CNOT gate with additional single qubit transformations, examples see
 \cite{kauffman8}. The unitary $\check{R}$-matrix (\ref{theorem}) is
 realized by
 \eq
\check{R}=\check R_-(\theta)|_{\theta=\varphi=0}=e^{i\frac \pi 4
(\sigma_x\otimes\sigma_y) }.
 \en

  In addition, with the unitary $\check{R}(\theta)$-matrix (\ref{bellrmatrix}),
  we have \eq \check{R}_\pm(\theta) \left(\begin{array}{c} |0 0\rangle \\| 0 1 \rangle \\
 |1 0\rangle \\ |1 1\rangle \end{array}\right)=
 \left(\begin{array}{l}
 \cos(\frac \pi 4-\theta) |0 0\rangle- e^{i\varphi}\sin(\frac \pi 4-\theta) |1 1\rangle
\\  \cos(\frac \pi 4-\theta) |0 1\rangle \mp \sin(\frac \pi 4-\theta) |1 0\rangle
 \\  \cos(\frac \pi 4-\theta)|1 0\rangle \pm  \sin(\frac \pi 4-\theta) |0 1\rangle)
\\  \cos(\frac \pi 4-\theta) |1 1\rangle + e^{-i\varphi} \sin(\frac \pi 4-\theta)
 |0 0\rangle \end{array}\right). \en Hence with the concept of the Bloch vectors on
 the Bloch sphere \cite{nielsen}, the variables $\theta$ and $\varphi$
 realize their geometric meanings and the construction of the CNOT  gate
 becomes clear.

To obtain other two-qubit quantum gates, for instance the CNOT
gate, we have to apply single qubit unitary transformations $A, B,
C, D$ which can be possibly found in the Bloch sphere
\cite{nielsen} by $SO(3)$ rotations, namely, \eq \label{universal}
 (A\otimes B)U_\pm(\theta)(C\otimes D)=P_\uparrow\otimes
1\!\! 1+ P_\downarrow\otimes\sigma_x=\textrm{CNOT}  \en which
yields the CNOT gate and where the states $|\uparrow\rangle$ and
$|\downarrow\rangle$ are the eigenvectors of $\sigma_z$,
$\sigma_z|\uparrow\rangle=|\uparrow\rangle,
\sigma_z|\downarrow\rangle=-|\downarrow\rangle$ and the projection
operators $P_\uparrow$ and $P_\downarrow$ have the forms
 \eq
 P_\uparrow=|\uparrow\rangle \langle
  \uparrow|, \qquad P_\downarrow=|\downarrow\rangle\langle
  \downarrow|.
 \en

 Define the $SO(3)$ rotation around the $\vec n$-axis by
 \eq D_{\vec n}(\theta)=e^{-\frac i 2 (\vec \sigma \cdot \vec n)\theta}   \en
 where $\vec \sigma=(\sigma_x, \sigma_y, \sigma_z)$. For examples:
 \eq  D_z(-\frac \varphi 2)=e^{i\frac \varphi 4 \sigma_z}, \,\,
  D_x(\frac \pi 2)=e^{-i\frac \pi 4 \sigma_x}, \,\, D_y(\frac \pi 2)
  =e^{-i\frac \pi 4 \sigma_y} \en
 satisfy
 \eq
 D_x(\frac \pi 2)D_z(-\frac \varphi 2)\sigma_{n_1}D_z(\frac \varphi
 2 ) D_x(-\frac \pi 2)=\sigma_z, \qquad D_z(-\frac \varphi 2)
 \sigma_{n_2}D_z(\frac \varphi 2)=\sigma_x. \en

Consider the time-evolution operator $U_+(\theta)$
(\ref{evolution1}). Choosing suitable single qubit
transformations, we obtain \eq ( D_x(\frac
 \pi 2)D_z(-\frac \varphi 2)\otimes D_z(-\frac \varphi 2))U_+
 (\theta)(D_z(\frac \varphi 2)D_x(-\frac \pi 2)\otimes D_z(\frac
 \varphi 2) )= e^{-\frac i 2 (\sigma_z\otimes\sigma_x)\theta}\en
which has another form \eq e^{-\frac i 2
(\sigma_z\otimes\sigma_x)\theta}=P_\uparrow\otimes e^{-\frac i 2
\sigma_x\theta} \, + \, P_\downarrow\otimes e^{\frac i 2
\sigma_x\theta}. \en Set $\theta=\frac \pi 2$. To construct the
CNOT gate, we need additional single qubit transformations
 \eq (\delta\otimes e^{i\frac \pi 4\sigma_x})
 e^{-i\frac \pi 4 (\sigma_z\otimes\sigma_x)}=\textrm{CNOT}
 \en in which the phase gate $\delta$ has the form
 $\delta=P_\uparrow-i\,P_\downarrow$, see (\ref{delta}).

Consider the time-evolution operator $U_-(\theta=-\frac \pi 2,
\varphi=0)$, namely the unitary $\check{R}$-matrix
(\ref{theorem}) given by $e^{i\frac \pi 4
(\sigma_x\otimes\sigma_y)}$ which is transformed into $e^{i\frac
\pi 4 (\sigma_z\otimes\sigma_x)}$ by \eq (D_y(-\frac \pi 2)\otimes
D_z(-\frac \pi 2))e^{i\frac \pi 4
(\sigma_x\otimes\sigma_y)}(D_y(\frac \pi 2)\otimes D_z(\frac \pi
2))=e^{i\frac \pi 4 (\sigma_z\otimes\sigma_x)}.\en So we obtain
another proof for {\bf Theorem  1} \cite{kauffman8}.

 \section{Concluding remarks}

 Motivated by the observation that there are certain natural similarities
 between quantum entanglements and topological entanglements, we derive
 the unitary solutions of the QYBE or the unitary $\check{R}(x)$-matrices
 via Yang--Baxterization and construct the related Hamiltonians for the
 standard and non-standard representations of the six-vertex model and the
 complete solutions of the non-vanishing eight-vertex model. With the Brylinksis'
 theorem \cite{BB}, the unitary $\check{R}(x)$-matrix is also a universal
 quantum gate except very special cases.

 The remark has to be made on the classification of the unitary
 $\check{R}(x)$-matrices we obtained. We classify them according to Yang--Baxterization.
 We don't try to classify them with other approaches since we can't obtain the
 complete solutions of the QYBE (\ref{qybe}) only via Yang--Baxterization. But
 the unitary solutions of the Yang--Baxter equation (the braid relation)
 have been classified \cite{dye}. For example, the first type of
 solution for the eight-vertex model in our case belongs the fourth family specified
 by \cite{dye}. But for the eight-vertex model, we have three types solutions of the
 BGR (\ref{bgr}) but have four types of the unitary $\check{R}(x)$-matrices.

 It is worthwhile arguing again that quantum entanglements for
 quantum information processing are related to not only topological
 entanglements but also geometric invariants. The previous one leads us to viewing
 the unitary braiding operator as a quantum entanglement operator and a
 universal quantum gate, while the latter one suggests us to regard
 unitary $\check{R}(x)$-matrices as universal quantum gates.
 In addition, it is important to mention another view of
 topological issues for quantum computing in terms of anyonic
 models, see \cite{freedman1, freedman2, freedman3, freedman4, freedman5, freedman6}.

\section*{Acknowledgements}

\indent  Y. Zhang is indebted to X.Y. Li for his constant
encouragement, guidance and support during the research. He thanks
the hospitality of Institut des Hautes {\'E}tudes Scientifiques,
Bures-sur-Yvette during the stay and the hospitality of Institut
f{\" u}r Theoretical Physik, Leipzig University during the stay.
This work is in part supported by NSFC--10447134.

For L.H. Kauffman, most of this effort was sponsored by the
Defense Advanced Research Projects Agency (DARPA) and Air Force
Research Laboratory, Air Force Materiel Command, USAF, under
agreement F30602-01-2-05022. The U.S. Government is authorized to
reproduce and distribute reprints for Government purposes
notwithstanding any copyright annotations thereon. The views and
conclusions contained herein are those of the authors and should
not be interpreted as necessarily representing the official
policies or endorsements, either expressed or implied, of the
Defense Advanced Research Projects Agency, the Air Force Research
Laboratory, or the U.S. Government. (Copyright 2004.) It gives
L.H. Kauffman great pleasure to acknowledge support from NSF Grant
DMS-0245588.

  \appendix

   \section{A practical revisit to Yang--Baxterization }

Yang--Baxterization \cite{jones} is a prescription deriving
solutions to the QYBE (\ref{qybe}) from the BGR (\ref{bgr}). We
consider a BGR $b$-matrix with two distinct non-vanishing
eigenvalues $\lambda_1$ and $\lambda_2$ taking the form \eq
b=\lambda_1\,P_1\,+\,\lambda_2\,P_2 \en where $P_1$ and $P_2$ are
the projection matrices satisfying \eq P_1+P_2=1\!\! 1, \qquad
P_1^2=P_1,\,\, P_2^2=P_2,\,\, P_1\,P_2=0. \en With the help of
Yang--Baxterization, the corresponding $\check{R}(x)$-matrix
modulo an overall scalar factor \cite{molin1, molin2} has the form
\eqa \label{baxterization1}
 \check{R}(x)&=&(\lambda_1+\lambda_2\,x )\, P_1 \,+\,(\lambda_2\,+\,\lambda_1\,x )\,
 P_2\nonumber\\
&=& b+ \lambda_1\,\lambda_2\,x\,b^{-1}
 \ena
where the inverse matrix $b^{-1}$ is given by \eq
 b^{-1}=\frac 1 \lambda_1\,P_1+\frac 1 \lambda_2\,P_2.\en

In the case that the BGR $b$-matrix has three distinct
non-vanishing eigenvalues $\lambda_1, \lambda_2$ and $\lambda_3$,
Yang--Baxterization \cite{molin1, molin2} leads to the following
formula \eq \label{baxterization2} \check{R}(x)= \lambda_1
\lambda_3 x(x-1)
b^{-1}+(\lambda_1+\lambda_2+\lambda_3+\lambda_1\lambda_3\lambda_2^{-1})
x 1\!\!1-(x-1)b. \en Changing the ordering of three different
eigenvalues in principle gives us different solutions of the QYBE
(\ref{qybe}). However, this formula is symmetric with respect to
interchanging $\lambda_1$ and $\lambda_3$, so usually we obtain
three types of the $\check{R}(x)$-matrices.

 With two formulas (\ref{baxterization1})
 and (\ref{baxterization2}), the $\check{R}(x)$-matrix is proportional to the
 unit matrix $1\!\! 1$ at $x=1$, namely,
\eq \check R(x=1)\propto 1\!\! 1 \en
  except the case of $\lambda_1+\lambda_2+\lambda_3+
 \lambda_1\lambda_3\lambda_2^{-1}=0$. Hence for simplicity, in the
 following we will not pay special attention to the case of $x=1$.

 One remark has to be made. The formula (\ref{baxterization1})
 has been proved to satisfy the QYBE (\ref{qybe}), however the formula
 (\ref{baxterization2}) does not have the general proof verifying that it is
 the solution of the QYBE (\ref{qybe}). Once the $\check{R}(x)$-matrix is obtained
 via Yang--Baxterization, it would be safest to check whether it truly
 satisfies the QYBE (\ref{qybe}).

 \subsection{Yang--Baxterization of the six-vertex model}

Consider a non-standard BGR $b$-matrix suitable for constructing
the Alexander polynomial \cite{kauffman0} \eq
b=\left(\begin{array}{cccc}
q & 0 & 0 & 0 \\
0 & 0 & 1 & 0 \\
0 & 1 & q-q^{-1} & 0 \\
0 & 0 & 0 & -q^{-1}
\end{array}\right)
\en where the deformation parameter $q$ has been assumed to be
non-vanishing. It has two distinct eigenvalues: $q$ and $-q^{-1}$.
In terms of the projection matrices $P_1(q)$ and $P_2(q)$: \eq
P_1(q)=\left(\begin{array}{cccc}
1 & 0 & 0 & 0 \\
0 & \frac 1 {1+q^2} & \frac q {1+q^2} & 0 \\
0 & \frac {q} {1+q^2} & \frac {q^2} {1+q^2} & 0 \\
0 & 0 & 0 & 0
\end{array}\right),\qquad
P_2(q)=\left(\begin{array}{cccc}
0 & 0 & 0 & 0 \\
0 & \frac { q^2} {1+q^2} & \frac {-q} {1+q^2} & 0 \\
0 & \frac {-q} {1+q^2} & \frac {1} {1+q^2} & 0 \\
0 & 0 & 0 & 1
\end{array}\right),
 \en
the BGR $b$-matrix is also given by \eq
b=q\,P_1(q)-q^{-1}\,P_2(q). \en With the help of
Yang--Baxterization, the BGR $b$-matrix corresponds to the
following $\check{R}(x)$-matrix satisfying the QYBE (\ref{qybe}),
\eqa \check{R}(x)&=&(q\,-\,q^{-1}\,x)\,P_1(q)\,+\,(-q^{-1}\,+\,
q\,x)\,P_2(q)  \nonumber\\
&=& b-x\, b^{-1}. \ena

\subsection{Yang--Baxterization of the eight-vertex model}

 The eight-vertex model assumes a general form
  \eq
 b=\left(\begin{array}{cccc}
 w_1 & 0 & 0 & w_7 \\
 0 & w_5 & w_3 & 0 \\
 0 & w_4 & w_6 & 0 \\
 w_8 & 0 & 0 & w_2
 \end{array}\right),\en
 in terms of  non-vanishing Boltzman weights $w_i$, $i =1, \cdots 8$.
 We will present the complete solutions of the BGR (\ref{bgr}) for the non-vanishing
 eight-vertex model and construct the corresponding $\check{R}(x)$-matrices via
 Yang--Baxterization.

 Introduce $f^{\kappa\lambda\omega}_{\alpha\beta\gamma}$ by \eq
 f^{\kappa\lambda\omega}_{\alpha\beta\gamma} =\sum_{\mu\nu\rho}
(b^{\mu\nu}_{\alpha\beta}b^{\rho\kappa}_{\nu\gamma}b^{\lambda\omega}_{\mu\rho}
-b^{\mu\nu}_{\beta\gamma}b^{\lambda\rho}_{\alpha\mu}b^{\omega\kappa}_{\rho\nu}),
 \en where indices take spin up or spin down, namely $\pm$ sign or spins $\pm \frac 1
 2$. With the above ansatz, we have $b^{\kappa\omega}_{\alpha\beta}=0$ for
 $\alpha+\beta\neq\kappa+\omega$ mod $2$. In the case of $\alpha+\beta+\gamma\neq\lambda+
 \omega+\kappa$ mod 2, the braid relation (\ref{bgr}) is satisfied
 automatically, $f^{\lambda\omega\kappa}_{\alpha\beta\gamma}=0$. Hence we have to
 treat thirty-two equations of the Boltzman weights.

 We solve all possible equations obtained by substituting
 the above ansatz of the eight-vertex model into the
 braid relation (\ref{bgr}). We observe the equation $(w_5-w_6) w_7
 w_8=0$ showing $w_5=w_6$. With the equations $(w_3-w_4)(w_1-w_5) w_8 = 0$ and
 $(w_3-w_4 ) (w_2-w_5 ) w_7 = 0$, we have to choose either $w_3=w_4, w_5=w_6$ or
 $w_3\neq w_4, w_5=w_1=w_2=w_6$.

Setting $w_5=w_1=w_2=w_6$ gives us $w_1^2=w_3^2=w_4^2$ and
$w_3^2+w_7 w_8=0$. In the case of $w_3\neq w_4$, we have
$w_3=-w_4$ and $w_1=\pm w_3$. The BGR $b_\pm$-matrices have the
forms \eq b_\pm=\left(\begin{array}{cccc}
w_1 & 0 & 0 & w_7 \\
0 & w_1 & \pm w_1 & 0 \\
0 & \mp w_1 & w_1 & 0 \\
-\frac {w_1^2} {w_7} & 0 & 0 & w_1
\end{array}\right)  \Longleftrightarrow
\left(\begin{array}{cccc}
1 & 0 & 0 & q \\
0 & 1 & \pm 1 & 0 \\
0 & \mp 1 & 1 & 0 \\
-q^{-1} & 0 & 0 & 1
\end{array}\right).\en
It has two eigenvalues $\lambda_1=1-i$, $\lambda_2=1+i$.  The
corresponding $\check{R}(x)$-matrices via Yang--Baxterization are
obtained to be \eq  \label{belltype} \check{R}_\pm(x)= b_\pm
\,+\,2\, x\, b_\pm^{-1}. \en

Imposing $w_5=w_6$ and $w_3=w_4$ in the given eight-vertex model,
the solutions of the QYBE (\ref{qybe}) have to satisfy the
following three independent equations \cite{molin2},
\eqa w_5^2-w_7 w_8 &=& 0, \nonumber\\
w_1^2-w_3^2-w_1 w_5+w_2 w_5 &=& 0, \nonumber\\
w_2^2-w_3^2+w_1 w_5-w_2 w_5 &=& 0. \ena

Since the $\check{R}(x)$-matrix is modulo an overall scalar
factor, we introduce  \eq q=\frac {w_7} {w_5 }, \qquad
q^{-1}=\frac {w_8} {w_5}, \qquad t=\frac {w_2} {w_5}, \qquad
z=\frac {w_3} {w_5} \en and rewrite the above equations in an
explicit way as
 \eqa t^2+(\frac {w_1} {w_5})^2 &=& 2 z^2 , \nonumber\\
(\frac {w_1} {w_5}-t)(\frac {w_1} {w_5}+t-2) &=& 0. \ena Therefore
we obtain two types of the BGR (\ref{bgr}). The first one is given
by taking $\frac {w_1} {w_5}=2-t$  and the second one is specified
by $\frac {w_1} {w_5}=t$.

In the first case of $\frac {w_1} {w_5}=2-t$, set
$z=(t^2-2t+2)^\frac 1 2$. The BGR $b_\pm$-matrices take the form
\eq b_{\pm}=\left(\begin{array}{cccc}
2-t & 0 & 0 & q \\
0 & 1 & \pm z & 0 \\
0 & \pm z & 1 & 0 \\
q^{-1} & 0 & 0 & t
\end{array}\right).
\en It has two distinct eigenvalues $1\pm z$ and so via
Yang--Baxterization the $\check{R}(x)$-matrices are obtained by
\eq \check{R}_{\pm}(x)=
  b_{\pm}+ x (1-z^2) b_{\pm}^{-1}
\en where  the matrix entries satisfy $w_1 w_2 + w_3 w_4= w_5 w_6
+ w_7 w_8$.

In the second case of $w_3=w_4$, namely take $\frac {w_1} {w_5}=t$
so that $z=\pm t$. The corresponding BGR $b_\pm$-matrix has the
form \eq \label{bgr4} b_{\pm}=\left(\begin{array}{cccc}
t & 0 & 0 & q \\
0 & 1 & \pm t & 0 \\
0 &\pm  t & 1 & 0 \\
q^{-1} & 0 & 0 & t
\end{array}\right).
\en It has three eigenvalues $1+t$, $1-t$ and $-1+t$. To apply the
formula (\ref{baxterization2}) for the case of three distinct
eigenvalues, we arrange three eigenvalues $\lambda_1$, $\lambda_2$
and $\lambda_3$ in three different orderings as follows \eq
\begin{array}{lccc}
\label{ordering}
& \lambda_1 & \lambda_2 & \lambda_3 \\
\textrm{the first ordering:} & 1+t & 1-t & t-1  \\
\textrm{the second ordering:}& 1+t & t-1 & 1-t \\
\textrm{the third ordering:}& 1-t & 1+t & t-1
\end{array}.
\en

For the first ordering and the second one, we have \eq
\lambda_1+\lambda_2+\lambda_3+\lambda_1\lambda_3\lambda_2^{-1}=0
\en so that the situation goes back to the case of two distinct
eigenvalues, \eqa
\check{R}_\pm(x)&=&-(x-1)(b_\pm \pm\,x\,(1-t^2)\,b_\pm^{-1}) \nonumber\\
&\propto& b_\pm \pm\,x\,(1-t^2)\,b_\pm^{-1} \ena where the plus is
for the first ordering and the minus is for the second ordering.
We only need discuss the case of the first ordering.

For the BGR  $b$-matrix (\ref{bgr4}) having three distinct
eigenvalues, with the first ordering, we have \eq
\check{R}_{\pm}(x) = b_\pm+ x (1-t^2) b_\pm^{-1};\en  with the
third ordering (\ref{ordering}), we have
 \eq \check{R}_{\pm}(x)= -(x-1)\lambda_2
  ( b_\pm-\lambda_1\lambda_3 x\,b_\pm^{-1})+(\lambda_2^2+\lambda_1\lambda_3) x
1\!\!1. \en

\end{document}